\documentclass[letterpaper]{article} 
\usepackage[preprint]{aaai2027}  
\usepackage[hyphens]{url}  
\usepackage{graphicx} 
\usepackage{natbib}  
\usepackage{caption} 
\usepackage{algorithm}
\usepackage{algorithmic}
\usepackage{enumitem}
\usepackage{amsmath}
\usepackage{amssymb}       
\usepackage{cite} 
\usepackage{alltt}
\usepackage{multirow}
\usepackage{tabularx}
\usepackage{xcolor}
\definecolor{cvprblue}{rgb}{0.21,0.49,0.74}
\usepackage[pagebackref,breaklinks,colorlinks,linkcolor=cvprblue,
citecolor=cvprblue]{hyperref}

\usepackage{newfloat}
\usepackage{listings}
\DeclareCaptionStyle{ruled}{labelfont=normalfont,labelsep=colon,strut=off} 
\floatstyle{ruled}
\newfloat{listing}{tb}{lst}{}
\floatname{listing}{Listing}

\lstdefinestyle{compact}{
  basicstyle=\ttfamily\scriptsize,
  breaklines=true,
  breakatwhitespace=false,
  columns=fullflexible,
  keepspaces=true,
  showstringspaces=false,
  frame=none,
  xleftmargin=0pt,
  aboveskip=2pt,
  belowskip=2pt,
}

\usepackage{booktabs}

\usepackage{hyperref}

\title{One Recipe, Many Harnesses: What Self-Evolution Encodes Across Languages and Models}
\author{
    Siqi Yang\textsuperscript{\rm 1,\rm 2}\footnote{Corresponding author: siqiyang@illinois.edu},
    Qianlan Yang\textsuperscript{\rm 1},
    Yu-Xiong Wang\textsuperscript{\rm 1},
    Saurabh Pujar\textsuperscript{\rm 2},
    Martin Hirzel\textsuperscript{\rm 2}
}

\affiliations{
    \textsuperscript{\rm 1}University of Illinois Urbana-Champaign
    \textsuperscript{\rm 2}IBM
}

\begin{document}

\maketitle

\begin{abstract}
Self-evolving harnesses are closed-loop systems in which an agent inspects its own rollouts and edits its prompts, tools, and memory. 
They reliably improve coding agents in evaluations, but prior work reports aggregate gains rather than analyzing what the evolved artifacts encode. 
It therefore remains unclear whether they encode benchmark-specific adaptations, language-specific engineering knowledge, or compensation for limitations of the underlying model.
We disentangle these factors by holding an evolution recipe fixed across a grid of eight programming languages (Multi-SWE-Bench) and three base models, and analyzing the resulting harnesses. 
The recipe routes every edit through a typed failure signal and records it as a falsifiable contract, making each modification attributable after evolution. 
Four findings emerge. 
(1)~The loop improves held-out solve rates over both a minimal seed and the manually designed mini-SWE-agent scaffold in most cells, but with two null regions.
(2)~Gains compensate recoverable execution defects, where defect mass is near zero, and gain is near zero; which defect dominates is cell-specific. A harness closes the gap between what a policy can do and what it does.
(3)~Evolved harnesses share an abstract playbook across languages but instantiate it with almost disjoint language ecosystem machinery.
(4)~The shared core transfers and can be distilled into one universal harness, while an ecosystem margin resists both and requires native re-evolution. 
Together, these results recast the evolved harness as a legible compensation layer, shaped jointly by the language's engineering demands and the model's behavioral gaps, rather than an opaque benchmark-tuned scaffold. Code is available at \href{https://github.com/IIDA-Institute/triage}{Github}
\end{abstract}


\section{Introduction}
\label{sec:intro}
Coding-agent progress on repository-level and long-horizon software engineering tasks has been driven by both stronger base models and better \emph{harnesses}---the editable prompts, tools, memory, and workflow scaffolding through which agents act~\citep{swebench, sweagent, wang2025openhands}.
As base models advance and application surfaces multiply, harness engineering has emerged as an important optimization subject: the same base model can improve substantially depending on the surrounding harness~\citep{ahe, xia2025live}. 
However, the optimal harness is model- and task-specific, and transfer behavior of such harnesses is insufficiently understood. 
This motivates a growing line of work on \emph{self-evolving harnesses}---outer-loop \emph{recipes} in which an evolving agent inspects inner-loop rollouts and edits the harness autonomously~\citep{hu2025automated, xia2025live, dgm, ahe}.

Recent work has established that self-evolving harnesses work: they beat both human-designed baselines and prompt-only self-improvement methods on standard evaluations~\citep{ahe, xia2025live, zhang2025agentic}. 
However, a more fundamental question is unanswered: what does a self-evolved harness actually encode? 
A harness that lifts pass@1 could reflect programming-language knowledge, compensation for the base model's cognitive limitations, or some combination of the two; it could also merely overfit the evaluated instances. Distinguishing the first two requires holding the evolution recipe fixed while varying the language and the model, and analyzing the resulting artifacts.
Without this design, it remains unclear when a harness will transfer, require re-evolution, or which gains are portable. 
We therefore apply a diagnostic-routed framework across a language $\times$ model grid, linking every edit to a specific failure signal.

We study programming language and model capability as the two axes within the code-fixing domain of Multi-SWE-Bench~\cite{multiswebench}.
Coding-agent research remains centered on Python benchmarks, while multilingual evaluations reveal substantial performance variation across languages~\citep{multiswebench,rashid_et_al_2025,baltaji_et_al_2025}. These gaps may partly reflect language-specific engineering demands that per-language harness evolution can address.
Beyond asking whether per-language evolution addresses these language-specific demands, using programming language as the second axis offers analytical advantages: language \emph{ecosystems}~(including build, test, and dependency tooling) impose concrete, comparable constraints, and fixing the code-repair domain reduces task-level variation. This makes language and model effects easier to distinguish.

We instantiate this design in a self-evolution framework built for mechanistic legibility, as shown in Fig.~\ref{fig:overview}.
An LLM analysis of failure trajectories emits typed signals that route each edit to a specific component and record it as a falsifiable contract, rather than relying on unstructured reflection or a scalar reward.
We introduce no new search algorithm; instead, holding the recipe fixed across all $8\times3$ settings lets us attribute structural differences to the (language, model) setting rather than to the search procedure.

Our empirical results support four claims that together reposition self-evolving harnesses from opaque scaffolds to legible compensation layers.

\textbf{(1) The recipe works, and where it does not is explicable.}
Held fixed across all $8\times3$ cells, the
loop improves held-out solve rates over the minimal seed and over the manually designed mini-SWE-agent scaffold~\citep{minisweagent} in most cells. Two regions are flat, one along
each axis: Python under every model, and GPT-5-mini in every language. Both are accounted for by (2) rather than left as exceptions.



\textbf{(2) Gains compensate recoverable execution defects}
The failures the loop repairs are process defects: editing a test file so that it collides with the hidden gold test patch, shipping source that does not build, submitting without running the target test. Which of these dominates is a property of the cell, not of the method---non-compiling changes account for 70\% of C++ rollout defects but under 10\% of Go's, so no single rule is the mechanism, and a static scaffold usually hard-codes one guess for all cells. The two nulls are the two ways that bound can bind: Python's base policies already follow the disciplines our detectors cover, and GPT-5-mini commits few such defects at all, leaving only localization and repair failures.


\textbf{(3) Evolved harnesses share a playbook but not its instantiation.}
Coding each harness into abstract
concepts and concrete language ecosystem markers reveals moderate cross-language overlap in concepts, but near-none in implementation: harnesses rediscover the same general disciplines yet express them through different commands, test paths, and layout conventions. Overall, 20-40\% of each harness is ecosystem-specific.


\textbf{(4) Reuse is limited by the ecosystem margin.}
The disciplinary core is genuinely portable: a harness transplanted unchanged onto a foreign language helps in 18 of 20 ordered pairs, and a distilled, language-agnostic harness matches native evolution on some ecosystems but not others. But it does not reach the native ceiling: transplants fall short in 14 of 20 pairs, and distillation recovers only 48-68\% of native gains on Java, C++, and TypeScript. Three measurements of different kinds — textual, distillation-based, and transplantation-based — converge on the same range. Native re-evolution thus remains necessary for complex ecosystems.



Together, these findings quantify the compensation view: the language determines which disciplines a task requires, model capability determines how many of them the base policy already supplies, and evolution installs the difference. 
Prior work showed that self-evolving harnesses outperform manual designs; we characterize what those gains consist of, when they transfer.

This paper makes the following contributions:

\noindent\textbf{Framing.} We identify programming language and
    model capability as two independently variable axes along which harness content and performance vary systematically.
   
\noindent\textbf{Instrumented method.} We introduce a diagnostic-routed evolution framework that links every edit to a falsifiable failure signal. Holding the recipe fixed makes differences legible and attributable to language and model.
    

\noindent\textbf{A structural and mechanistic account of harness gains.} Across the grid, gains compensate recoverable execution defects, and their size tracks the defect mass a cell exhibits. Which defect binds shifts systematically with the language, so no single rule is the mechanism and a static scaffold must hard-code one guess for all cells. The account also covers its two null regions (Python and GPT-5-mini) and locates 20-40\% of each harness in ecosystem-specific machinery.

\noindent\textbf{A transfer study.}
We quantify how far evolved harnesses port: the disciplinary core survives both transplantation onto foreign languages (positive in 18 of 20 ordered pairs) and distillation into a single universal harness, but how much survives depends sharply on the target, leaving an ecosystem margin that neither recovers. This yields practical guidance on when to reuse, distill, or re-evolve a harness.

\section{Related Work}

\subsection{Coding Agent Benchmarks and Harnesses}
Repository-level code repair has become a prominent testbed for coding agents because it combines reasoning, planning, tool use, and long-horizon interaction~\citep{swebench}. Performance depends not only on the model but also on the \emph{harness}: changing the agent--computer interface or scaffold while holding the model fixed can substantially affect solve rates~\citep{sweagent,wang2025openhands}. Yet leaderboard systems often use different, sometimes unreleased, harnesses, obscuring whether gains come from the model or scaffold. Harness effects may also interact with model choice, but this coupling remains under-studied. The mini-SWE-agent harness is small yet achieves competitive performance~\citep{minisweagent}.
Multi-SWE-bench holds the code-repair domain fixed while spanning multiple programming languages, making language a useful analytical axis~\citep{multiswebench}. Existing multilingual evaluations already adapt agents to new ecosystems: Both Multi-SWE-bench and SWE-PolyBench ship per-language agent variants~\citep{multiswebench,rashid_et_al_2025}. These efforts show that multilingual evaluation often involves language-specific harness changes, but treat those changes as engineering accommodations rather than subjects of systematic analysis.

\subsection{Self-Evolving Harness Systems}
Self-evolving harness systems use an LLM to propose harness edits, evaluate them on agent rollouts, and retain beneficial variants, but differ in what they evolve and how edits are selected. DGM evolves the agent repository through open-ended archive search~\citep{dgm}; AHE exposes seven editable component types and binds each edit to a falsifiable prediction~\citep{ahe}; HarnessX organizes harnesses along nine composable dimensions~\citep{chen2026harnessx}; HarnessFix diagnoses failures within a seven-layer harness taxonomy before applying scoped repairs~\citep{harnessfix}; and
HGM uses descendant performance
to guide search~\citep{wang2026huxleygodel}.
Related methods evolve cumulative context or skill artifacts. ACE incrementally updates an evolving playbook~\citep{zhang2025agentic}, SkillOpt applies held-out-gated edits to a skill document~\citep{yang2026skillopt}, and MUSE-Autoskill creates and refines reusable skills from experience~\citep{museautoskill}. 
Across these systems, an important design choice is how failures are translated into edits: methods use layer attribution, predicted task-level effects, rollout trajectory diagnosis, or held-out validation rather than relying only on unconstrained reflection.
Prior work varies substantially in benchmarks, models, editable components, and search procedures, making cross-setting differences difficult to attribute. We instead hold both the evolution recipe and routing scheme fixed across a language $\times$ model grid to study how grid cells affect the resulting harnesses.

\subsection{Externalized Tuning and Joint Adaptation}
\label{sec:related:posttrain}
A related line optimizes textual and program-level artifacts. DSPy represents LM pipelines as declarative programs~\citep{khattab_et_al_2024}; OPRO uses candidate--score histories to improve prompts~\citep{opro}; TextGrad propagates textual feedback through compound systems~\citep{textgrad};
AutoPDL uses feedback to decide whether to use an agent and
how to configure it~\citep{spiess_et_al_2025};
and GEPA combines reflection with evolutionary selection, outperforming GRPO in its reported evaluations with fewer rollouts~\citep{agrawal2025gepa}.
Other work adapts harnesses and model parameters jointly. SIA updates both the scaffold and model weights~\citep{SIA}, while HarnessForge co-evolves harnesses with harness-conditioned policy adapters~\citep{harnessforge}. These methods primarily evaluate optimized artifacts through task performance.

\section{\textsc{TRIAGE}: Harness Self-Evolution}
\label{sec:method}

\begin{figure}
\centerline{\includegraphics[width=\columnwidth]{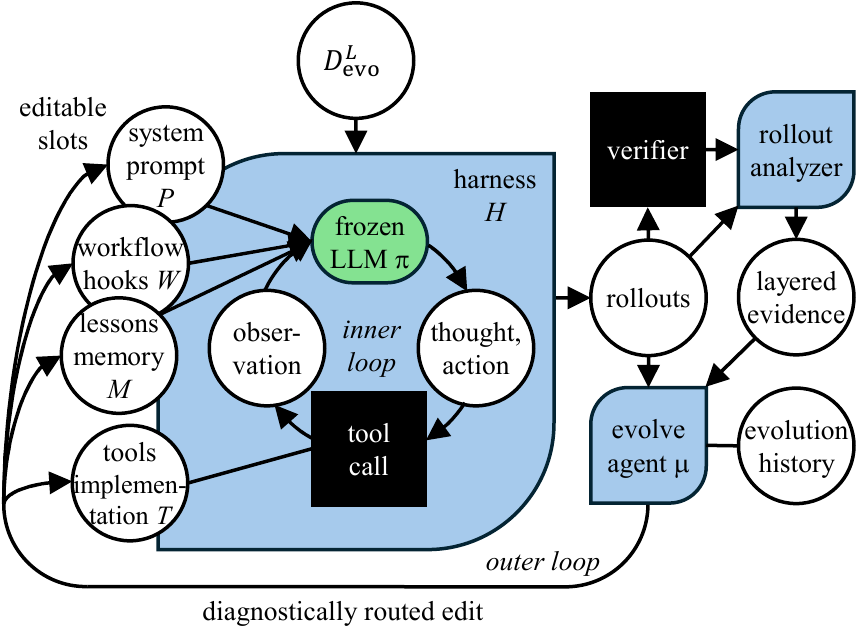}}
\caption{\label{fig:overview}Overview of harness self-evolution recipe.}
\end{figure}

Fig.~\ref{fig:overview} presents \textsc{TRIAGE}
(\textbf{T}yped \textbf{R}outing and \textbf{I}nstrumented \textbf{A}ttribution for \textbf{G}uided \textbf{E}volution), the instrumented evolution loop we hold fixed across the grid. We design TRIAGE for legibility: typed routing and contract-level records leave every edit attributable to a specific failure signal after evolution, which enables the harness-content analyses of Sec.~\ref{sec:exp:specificity} to Sec.~\ref{sec:exp:transfer}. We build on AHE's observability-driven backbone~\citep{ahe}, retaining its frozen policy, explicit editable harness, layered evidence, and later-round validation. \textsc{TRIAGE} adds two mechanisms: instrumented attribution incorporates verifier- and budget-side evidence, and typed routing uses the resulting diagnosis to prioritize where the harness changes.

For a language ecosystem \(L\), frozen task policy \(\pi\), and outer-loop
driver \(\mu\), the finite-round output is
\begin{equation}
H_{L,\pi,\mu}^{(R)}
=
\operatorname{Evolve}^{R}
\!\left(H_0,D_{\mathrm{evo}}^{L};\pi,\mu\right).
\label{eq:finite-evolution}
\end{equation}
This is the checkpoint returned after \(R\) rounds, rather than a global
optimum. A disjoint \(D_{\mathrm{test}}^{L}\) is reserved for final
evaluation and never used to propose or select edits. In controlled
comparisons, \(\mu\) and the evolution protocol are fixed.

\subsection{Modular Harness Evolution}
\label{sec:method:loop}

The harness exposes four versioned, file-level slots,
\mbox{\(H=(P,W,M,T)\)}: the system prompt \(P\), workflow hooks~\(W\), lessons memory~\(M\), and tool
implementations \(T\). These respectively encode persistent behavior, state-dependent guidance, reusable cross-instance lessons, and executable capabilities. We implement them as a lightweight plugin layer on mini-SWE-agent~\citep{minisweagent}; the abstraction requires only inspectable control surfaces around a frozen policy. Skills, sub-agents, or multi-agent orchestration are out of scope.

Every setting starts from the same minimal seed \(H_0\): a short prompt, one shell tool, and empty hooks and memory. Memory is capacity-bounded and admits only rules marked as cross-instance reusable, so the cap pressures distillation rather than accumulation; detailed component constraints are given in Appendix A.

At round \(r\), the inner loop runs \(H^{(r)}\) on
\(D_{\mathrm{evo}}^{L}\), producing rollout trajectories and verifier outcomes. The rollout analyzer combines these artifacts to form layered evidence. Conditioned on this evidence, the current harness, and the evolution history, the evolve agent~$\mu$ proposes evidence-backed edits and predicts their task-level improvements and regressions. The next round tests those predictions on the evolution split. We use one linear checkpoint sequence, retaining supported updates and rejecting or reverting regressive ones, without population search or branching. The fixed seed, interface, evidence schema, budget, and selection procedure make the returned harnesses comparable across settings.

\subsection{Failure Attribution and Typed Routing}
\label{sec:method:obs}
\label{sec:method:route}

A trajectory records what the task agent attempted, while a scalar verifier outcome records whether the submission passed. Neither alone explains why it failed, particularly when the agent itself believes it succeeded. \textsc{TRIAGE} therefore instruments two complementary evidence channels. Verifier-side evidence includes the scored submission, resulting environment state, and stage-level execution record, including stages that did not run. For example, it can expose interference between a submitted patch and a held-out test patch that appears in the transcript only as an ordinary failure. Budget-side evidence includes termination reasons and resource use, distinguishing a ceiling-interrupted attempt from a completed but incorrect submission.

The analyzer represents these records using a fixed, typed attribution schema. Its broad classes comprise verifier interaction, verification, localization, incorrect repair, and efficiency; exact signals appear in Appendix A. These types organize observations but prescribe neither a semantic cause nor a remedy: the analyzer infers the cause from the complete evidence. Routing is advisory. The attribution reaches $\mu$ as a dossier---leading cause plus failing stages, repository state, budget trace, and competing signals---not a verdict; $\mu$~weighs it and owns the causal claim. Routing therefore prioritises the search rather than fencing it. Persistent guidance typically enters the prompt or memory, state-dependent intervention a hook, and executable support a tool. Because the diagnosis is advisory, a misattribution costs an iteration in the typical case: the resulting edit tends to fail its own prediction and be reverted.

Following AHE's decision-observability contract~\citep{ahe}, attribution is treated as a falsifiable hypothesis rather than an oracle. Later rollouts compare predicted task effects with observed outcomes and reject or revert unsupported changes. This validation is checkpoint-level. Detailed schemas, routes, selection criteria, and examples are deferred to Appendix A. Beyond our coding-agent instantiation, \textsc{TRIAGE}'s requirements---a frozen policy, an inspectable harness, structured execution records, and measurable resource use---are not specific to code repair, though we evaluate only that setting.

\section{Experiments}
\label{sec:exp}

We evaluate the self-evolving harness loop of Sec.~\ref{sec:method} on Multi-SWE-Bench~\cite{multiswebench} across eight programming languages $L\in\{$\texttt{cpp}, \texttt{c}, \texttt{java}, \texttt{rust}, \texttt{typescript}, \texttt{javascript}, \texttt{go}, \texttt{python}$\}$ and three rollout models $\pi\in\{$Claude~Haiku~4.5, GPT-5-mini, DeepSeek-V4-Flash$\}$, forming an $8 \times 3$ grid of \mbox{$\langle L,\pi\rangle$} cells. We evaluate four claims, corresponding to the four results of Sec.~\ref{sec:intro}:

\begin{itemize}[leftmargin=*]

\item[\textbf{(C1)}] \textbf{Effectiveness.} 
The loop yields held-out gains over both a minimal seed~$H_0$ and mini-SWE-agent across the majority of \mbox{$\langle L,\pi\rangle$} cells, with two null regions---one per axis of the grid---that the mechanism of C2 accounts for (Sec.~\ref{sec:exp:main}).

\item[\textbf{(C2)}] \textbf{Compensation Mechanism and Its Ceiling.} 
Gains come from compensating \emph{recoverable execution defects}: the gap between what a base policy~$\pi$ can do and what it does under a bare scaffold. The harness closes this gap without raising the capability ceiling (Sec.~\ref{sec:exp:mechanism}).

\item[\textbf{(C3)}] \textbf{Cell-Specificity.} 
Because \emph{which} defect dominates differs by cell, evolved harnesses converge on abstract concepts while diverging in concrete instantiation, with $20$--$40\%$ of each harness bound to its target ecosystem (Sec.~\ref{sec:exp:specificity}).

\item[\textbf{(C4)}] \textbf{Bounded Portability.} 
The shared disciplinary core distills into one universal harness and transfers across languages, while an ecosystem margin limits both---triangulated by three independent measurements (Sec.~\ref{sec:exp:universal}--\ref{sec:exp:transfer}).

\end{itemize}

\subsection{Experimental Setup}
\label{sec:exp:setup}

\paragraph{Benchmark and Splits.}
We instantiate the 8 language-specific subsets of Multi-SWE-Bench~\citep{multiswebench}.
For each language~$L$, we randomly sample 20 instances for the evolution split~$D_\textrm{evo}^L$ and hold out 50 instances~$D_\textrm{test}^L$ for testing, yielding two disjoint sets $D_\textrm{evo}^L \cap D_\textrm{test}^L = \emptyset$. 

\paragraph{Models.}
For each model configuration, the task policy $\pi$ and outer-loop driver $\mu$ are instantiated by the same underlying model. Cross-model comparisons therefore reflect their joint effect. We evaluate on $n_{\text{model}} = 3$ models~$\pi$ spanning distinct capability profiles: two closed-weight models (Claude Haiku 4.5 and GPT-5-mini) and one open-weight model (DeepSeek-V4-Flash).

\paragraph{Compared Harnesses.}
Under identical steps, cost, and rollout budgets, we compare three harnesses per cell:
\begin{enumerate}[label=(\roman*)]
    \item \emph{Minimal Seed} ($H_0$, lower bound): 
    a stripped-down ReAct loop with a single bash tool, retaining only task descriptions and commit protocols without memory or hooks.
    \item \emph{Mini-SWE-Agent}~\citep{minisweagent}: A widely adopted, manually designed static scaffold.
    \item \emph{Evolved Harness} ($H^{(R)}_{L,\pi,\mu}$): The output of our self-evolution loop after $R=3$ rounds using~$D_\textrm{evo}^L$.
\end{enumerate}

\paragraph{Metrics and Protocol.}
Unless otherwise stated, our primary metric is \textsc{mean\_solve@3}: the mean solve rate across $k=3$ independent rollouts per held-out instance in~$D_\textrm{test}^L$.

\subsection{Main Results: Held-Out Gains}
\label{sec:exp:main}

\begin{figure*}[t]
\centering
\includegraphics[width=1.00\linewidth]{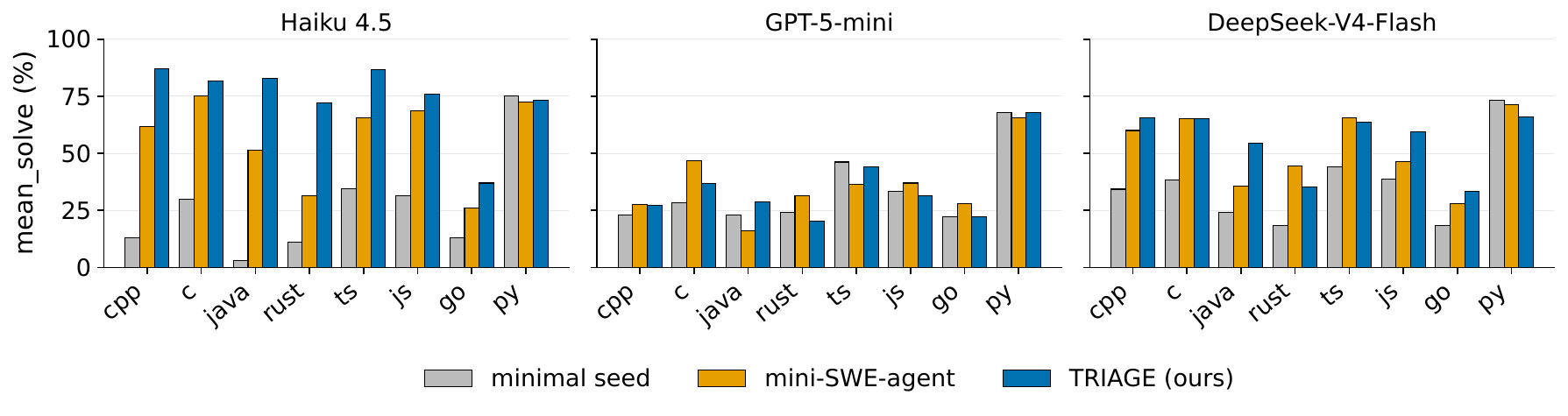}
\caption{\textbf{Main results}. Performance comparison between minimal seed, \texttt{mini-SWE-agent}, and evolved harnesses.}
\label{fig:main}
\end{figure*}

Fig.~\ref{fig:main} presents held-out results across all \mbox{$8\times3$ cells}. The evolved harness~$H^{(R)}_{L,\pi,\mu}$ improves over the minimal seed~$H_0$ in most cells and matches or exceeds the manually designed \mbox{mini-SWE-agent} in 14 of 24 cells; on Claude Haiku~4.5, the ordering \mbox{($H^{(R)}_{L,\pi,\mu}>\textrm{mini-SWE-agent}>H_0$)} is strict on every non-Python language.

Two exceptions fall along the two axes of the grid. Along the language axis, Python gains $\approx\!0$ under all three models; along the model axis, GPT-5-mini gains little in any language. The difference lies within the rollout standard deviation (Appendix C), so we read them as null results rather than regressions. Both nulls have the mechanism we identify in Sec.~\ref{sec:exp:mechanism}.

\subsection{Mechanism: What the Harness Compensates}
\label{sec:exp:mechanism}

What does the outer loop repair? Among failures to converted successes, many are \emph{model-side execution defects}, not its understanding of the bug.
We call a defect \emph{harness-recoverable} when (i) the rollout would plausibly have succeeded had the agent behaved differently at a step it was already capable of executing, and (ii) the corrective can be stated as a rule the harness installs ex ante. 
Three such defects are frequent enough to detect automatically: \textbf{editing a test file}, which collides with the hidden gold test patch (`git apply' fails) and zeroes an otherwise-correct source fix (\emph{matched discipline:} ``never modify test files''); \textbf{shipping non-compiling source}, so the test executable is never produced (``verify it compiles before submitting''); and \textbf{submitting without running the target test}, so wrong or incomplete fixes go undetected (``run the failing test before submitting'').

These are instances of the class, not a taxonomy of it: the loop also writes lower-frequency, cell-idiosyncratic rules such as repository layout conventions, per-ecosystem invocation forms, and diff-scope constraints. Our claim is accordingly directional: compensating recoverable execution defects is a principal source of the gains we observe. We do \emph{not} claim that measured defect mass upper-bounds the gain
available in a cell; our detectors cover only part of the recoverable space.

\begin{figure}[t]\centering
\includegraphics[width=1.00\linewidth]{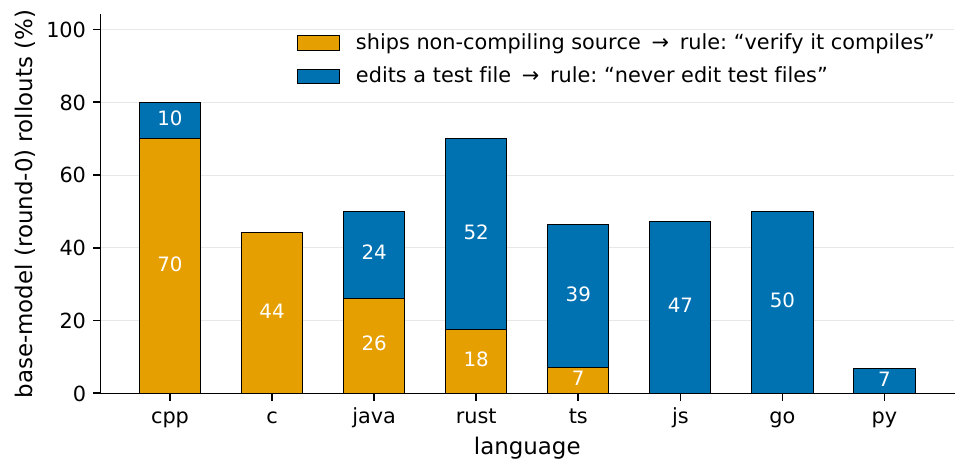}
\caption{\textbf{Composition of automatically detected execution defects by language.} 
Bars show the \emph{relative mix} of the two detectors that fire often: shipping non-compiling source vs.\ editing test files that collide with the hidden gold test patch. The mix shifts from build failures on compiled languages~(C++, C) to test-file edits elsewhere. 
}
\label{fig:selfsab}
\end{figure}

\paragraph{Which defect dominates is cell-specific.} Among the defects our detectors cover, composition shifts sharply across languages (Fig.~\ref{fig:selfsab}). 
On compiled languages the dominant mode is a non-compiling change---C++ $70\%$, C $44\%$---while
test-file editing is rare ($\le\!10\%$); on the scripting and systems targets the reverse holds. No single rule is therefore
``the'' mechanism. This is the operational argument for per-cell evolution: a harness must be told \emph{which} defect its cell exhibits.

\paragraph{Two null regions, consistent with this account.} Both nulls were identified post hoc; we read them as consistent with the mechanism rather than as tests of it. They arrive by different routes, one along each axis. \emph{Python (language axis):} for all three rollout models, Python's detector-covered defect rate is the lowest of the eight languages ($5.6\%$), so the effect tracks the language rather than any one model's competence---the rules the loop would install are ones these policies already follow.
Python is the one region where evolution buys nothing: the point estimate lies within the rollout standard deviation, so we claim only the absence of benefit, not a regression. 
\emph{GPT-5-mini (model axis):} our detectors fire seldom, and the loop reaches the same diagnosis independently---the \texttt{memory.md} it evolves contains no test-file rule at all, on languages where every other model's harness installs one as its first entry. What remains are failures of localizing the defective code and synthesizing a correct fix, which no rule addresses. We do not view GPT-5-mini as a stronger model; in fact, under the same mini-swe-agent, GPT-5-mini is the weakest. This suggests that models possess qualitative properties that are independent of their quantitative capability level.

\paragraph{Scope: compensation, not a raised ceiling.} Within the failure modes we can detect, the gain is compensation rather than capability extension: we find no recovered instance in which the evolved harness enables a repair the base policy could not otherwise express. We do not claim this holds outside the detected classes. Two practical readings follow. First, a base policy's measured defect profile is a cheap leading indicator of where harness engineering pays off---though it is an indicator, not an estimate of headroom, since the recoverable space is wider. Second, within the regime where a harness helps at all, an automatically evolved one outperforms a manually designed scaffold (Sec.~\ref{sec:exp:main}), precisely because the binding defect differs per cell and a static scaffold must hard-code one guess for all of them.


\subsection{Evolved Harnesses Are Cell-Specific}
\label{sec:exp:specificity}

The main results ship a separate harness per (language, model) cell. Are these genuinely distinct artifacts, or one generic playbook copied around? We answer by inspecting their content. We parse each evolved harness into codeable units and tag every unit at two levels: the abstract concept it expresses (e.g.\ ``localize from the test output'', ``minimal diff'') and any concrete ecosystem marker it names (a build tool, a test-path glob); a unit is \emph{ecosystem-specific} if it names such a marker. Across the evolved language cells, we then measure cross-language Jaccard overlap at each level, for the two rollout models with full coverage. We restrict the structural analysis to the 14 cells where the loop produces a held-out gain. The excluded cells cannot speak to what a working harness encodes (Sec.~\ref{sec:exp:mechanism}).

\begin{figure}[t]\centering
\includegraphics[width=1.00\linewidth]{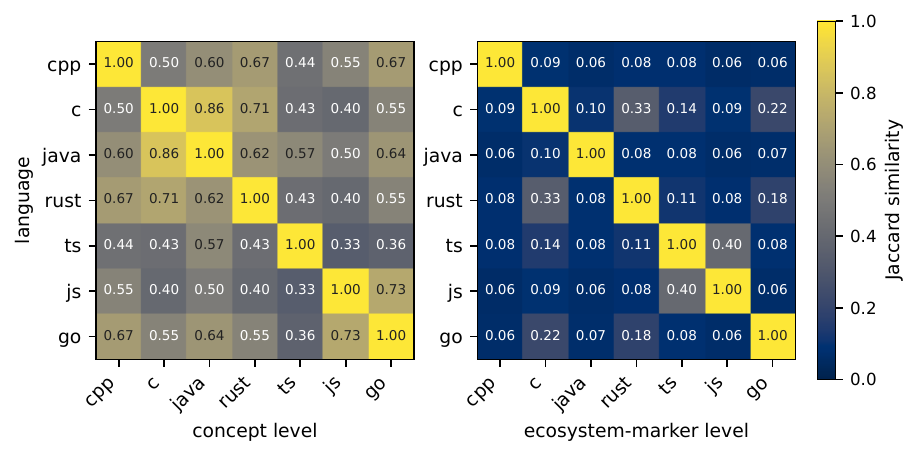}
\caption{Cross-language harness overlap for Claude Haiku 4.5 across the seven evolved languages. Left column: abstract-concept Jaccard---moderate overlap (mean $0.55$), a shared generic playbook. Right column: concrete ecosystem-marker Jaccard---near-zero overlap (mean $0.12$), the specific build/test/layout tokens each harness names barely co-occur across languages.}
\label{fig:c2jaccardhaiku}
\end{figure}

\paragraph{Concepts partly converge; instantiations do not.} Harnesses evolved for different languages overlap substantially in \emph{what} they aim to enforce and almost not at all in \emph{how} they enforce it. At the concept level, mean off-diagonal Jaccard is $0.55$ for Haiku (Fig.~\ref{fig:c2jaccardhaiku}) and $0.57$ for DeepSeek (See Appendix C): a sizable shared core of general disciplines, though far from identical---each language also elicits concepts the others do not. At the ecosystem-marker level, the overlap falls by roughly a factor of four, to $0.12$ and $0.14$: the build commands, test-path globs, and header layouts each harness actually names barely co-occur. 

The concept-level overlap is high but not near-unity, for two reasons we do not attempt to separate: languages genuinely differ in which disciplines matter (compiled targets
elicit build-verification concepts that interpreted ones never raise); and our concept tags come from an LLM rubric with imperfect granularity, which splits near-synonymous concepts and deflates overlap. We therefore rest the claim on the \emph{gap between the two levels} rather than on the absolute value at either: an evolved harness is not a copy of a universal artifact but a partly shared playbook instantiated on the concrete surface of its ecosystem---the structural form of C2, and the means by which the uniform gains of C1 are delivered by individually specialized harnesses.

\begin{figure}[t]\centering
\includegraphics[width=1.00\linewidth]{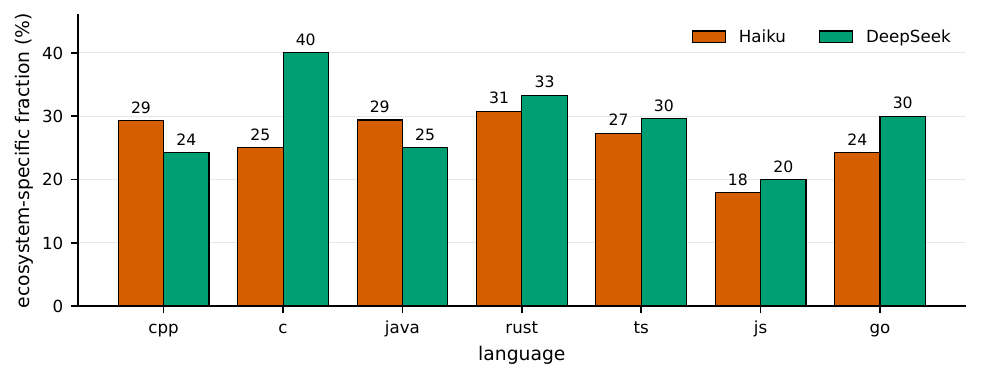}
\caption{Ecosystem-specific fraction of each evolved harness, per language, for Haiku and DeepSeek.}
\label{fig:c2ecofrac}
\end{figure}

Fig.~\ref{fig:c2ecofrac} sizes the divergent part directly: every evolved harness is $20$--$40\%$ ecosystem-specific (mean $0.26$ Haiku, $0.29$ DeepSeek), consistent across two rollout models. A fifth to two-fifths of what the loop writes is therefore unusable outside its target ecosystem.

\subsection{Universal Distillation: Isolating the Portable Component}
\label{sec:exp:universal}

Sec.~\ref{sec:exp:specificity} found that harnesses share concepts but not instantiations. If that split is functional, stripping the instantiations should preserve the shared part and lose the rest. We test this directly: using the meta-model, we aggregate the evolved memory rules of all languages into ten generic guidelines. The resulting \emph{universal harness} is memory-only and names no build command, test path, or file convention. We deploy it unchanged on DeepSeek's held-out set and report \emph{retention}
$\rho=(u-m)/(n-m)$, the fraction of that language's native gain it recovers, where $m$ is
the minimal-seed score, $n$ is the natively-evolved score, and $u$ is the transferred score.
\begin{figure}[t]\centering
\includegraphics[width=1.0\linewidth]{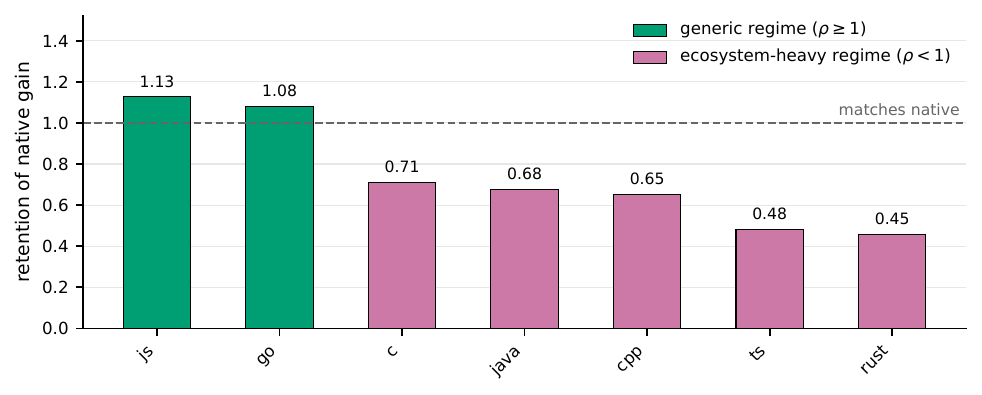}
\caption{\textbf{Universal harness vs.\ native evolution.} Fractional retention $\rho$ of
each language's native gain by a single language-agnostic, memory-only harness, across the
seven evolved languages.}
\label{fig:universal}
\end{figure}

\paragraph{Two regimes.} 
Fig.~\ref{fig:universal} separates the seven evolved languages---the same seven characterised textually in Sec.~\ref{sec:exp:specificity}. On \emph{generic} targets---Go ($1.08$), JavaScript ($1.13$)---the stripped harness is statistically indistinguishable from native: where the gain is purely disciplinary, per-language specialization adds nothing. On \emph{ecosystem-heavy} targets---Java~($0.68$), C++ ($0.65$), TypeScript ($0.48$)---it recovers only half to two-thirds. The residual $32$--$52\%$ is an \emph{ecosystem margin}: gain that exists only when the harness names the target's own build tools, test runners, and layout conventions. Its magnitude agrees with the textual measurement of Sec.~\ref{sec:exp:specificity}, where $20$--$40\%$ of each evolved harness is ecosystem-specific by content---the same quantity, measured once in text and once in held-out solve rate.

\subsection{Cross-Language Transfer is Structurally Bounded}
\label{sec:exp:transfer}

Distillation removes the ecosystem component; transplanting a \emph{foreign} harness keeps one, just the wrong language's. The two therefore bracket the native ceiling from opposite sides. We deploy the harness evolved natively on source $A$, unchanged, on the held-out set of target $B$, and report the same retention $\rho_{A\to B}$; the rollout model is DeepSeek-V4-Flash in every cell, so differences isolate the harness. We use the five languages spanning the two regimes just identified---generic (Go, JS) and ecosystem-heavy (C++, Java, TS)---omitting C (redundant with the other compiled targets), Rust (small native gain makes $\rho$ unstable), and Python (native gain $\le\!0$, so $\rho$ is undefined).

\begin{figure}[t]\centering
\includegraphics[width=0.9\linewidth]{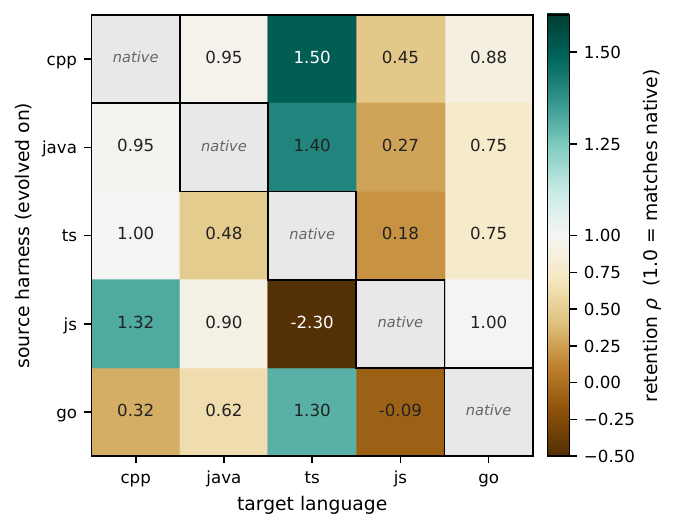}
\caption{\textbf{Cross-language transfer retention $\rho_{A\to B}$.} Rows: the source
harness, evolved natively on $A$. Columns: the target it is deployed on. The grey diagonal
is the native harness.}
\label{fig:xfer5}
\end{figure}

\paragraph{Most of the discipline travels.} 
Transfer is positive in $18/20$ off-diagonal cells, with mean retention $0.63$ ($0.79$ excluding the single strongly negative cell): a harness carried over from an unrelated language recovers a substantial share of the target's gain with no adaptation at all. This is the behavioural counterpart of the shared concept-level core, and it is what makes evolved harnesses reusable for bootstrapping a language the loop has never seen.

\paragraph{But native remains a ceiling.} 
Retention falls below $1$ in $14/20$ cells, and full recovery occurs in only $6/20$. The shortfall is not confined to weak sources: the strongest source, C++ (row mean $0.95$), reaches native on just one of its four targets. Roughly a third of what native evolution installs is thus supplied by none of the five sources we tested. Where full recovery \emph{does} occur, it is structured, not random---it concentrates where the source already carries the discipline the target demands, most visibly for compiled-language sources deployed on TypeScript (C++ $1.50$, Java $1.40$, Go $1.30$). Transfer completes when there is no language-specific gap left to bridge.

\paragraph{Exporting and importing are different properties.} 
The two marginals are nearly uncorrelated, as the mechanism of Sec.~\ref{sec:exp:mechanism} predicts: a harness exports the discipline matched to \emph{its} dominant defect, and a target is recoverable only if that is also the defect \emph{it} exhibits. JavaScript makes the point twice over. As a source it exports well into typed targets (C++ $1.32$, Go $1.00$, Java $0.90$); as a target it is the least recoverable language in the matrix (column mean $0.20$), with no foreign harness recovering even half its native gain.

\paragraph{The one wall: a discipline the ecosystem does not enforce for free.} 
The single strongly negative cell is JS$\to$TS ($-2.30$), and its asymmetry with JS$\to$C++ ($1.32$) identifies what transfer actually fails to carry. A JavaScript-evolved harness contains the language-general rule ``run the failing test before submitting'' but no compile-verification step, since interpreted JavaScript has nothing to compile. On C++ and Java, that omission is harmless: executing the target test \emph{entails} building the project, so the generic rule silently supplies compile verification and the gap never opens---which is why compiled targets are the easiest to import into. On TypeScript, the entailment breaks. Test runners routinely transpile without type-checking, so a patch can run its tests green while \texttt{tsc} rejects it, and the grader type-checks. Verification must therefore be named \emph{explicitly} in the harness or it does not happen at all, and a harness that never learned to name it ships patches that never build. The boundary a portable harness cannot cross is not interpreted-versus-typed as such, but the point at which a discipline stops being a free consequence of the ecosystem's own execution chain---exactly the compensation role of Sec.~\ref{sec:exp:mechanism}, seen from its failure side.

\paragraph{The two experiments agree.} 
On ecosystem-heavy targets, the foreign specialized harness recovers strictly more than the stripped universal one (C++ $0.90$ vs.\ $0.65$; Java $0.74$ vs.\ $0.68$) yet still falls short of native, giving $\text{universal}\le\text{transplant}<\text{native}$; on generic targets the order reverses (Go $0.84$ vs.\ $1.08$, JS $0.20$ vs.\ $1.13$), the foreign ecosystem baggage being dead weight or worse. The total gain therefore decomposes into a large portable disciplinary component and a remainder that neither reuse strategy recovers (C4): externalized harness engineering is a genuine reusable asset, but not a substitute for native evolution.

\section{Conclusion}

We studied what self-evolved harnesses encode by applying a fixed, instrumented evolution recipe across eight programming languages and three base models.
Typed failure signals and falsifiable edit contracts make the resulting harnesses traceable.
The evolved harness improves over both a minimal seed and mini-SWE-agent in most settings. Gains come from correcting recoverable execution defects. 
Python and \textsc{GPT-5-mini} form two null regions, in which the defects our detectors cover are uncommon, and evolution has little left to install.
Across settings, harnesses learn similar behavioral disciplines but express them through language-specific conventions.
This shared core is portable but incomplete: a distilled language-agnostic harness matches native evolution on generic targets yet recovers part of the native gain on ecosystem-heavy ones.
These results characterize the evolved harness as a compensation layer: it improves how a model applies its existing capabilities, with the required guidance shaped jointly by model behavior and language ecosystem.
Under testing conditions, existing harnesses appear useful starting points, but cannot always replace language-specific evolution.
Our evidence comes from repository-level repair under one evolution recipe with a single run per setting, so we read aggregate structure; whether the same decomposition into a portable disciplinary component and an ecosystem-bound remainder survives at longer horizons remains open.

\bibliography{aaai2027}

\clearpage
\appendix
\section{TRIAGE Implementation Details}
\label{app:triage}

This appendix specifies the parts of TRIAGE that Sec.~\ref{sec:method} summarises: the harness slots and
their admission constraints (Sec.\ref{app:slots}), the typed attribution schema
(Sec.~\ref{app:schema}), the routing table (Sec.~\ref{app:routing}), the edit-contract and selection
machinery (Sec.~\ref{app:contracts}), the meta-model roles and their inputs (Sec.~\ref{app:prompts}),
and one complete round traced end to end (Sec.~\ref{app:worked}).

\subsection{Harness Slot Specifications and Constraints}
\label{app:slots}

The harness $H=(P,W,M,T)$ is a directory of versioned files. Every slot is admitted through a
\emph{smoke test} before it can enter a round: a proposed edit that fails validation is rejected
and the round proceeds with the previous contents, so a malformed edit costs an iteration but
cannot corrupt the incumbent.

$P$ prompt: System and per-instance templates (plus observation and
format-error templates). Free-form text; unbounded in length. \\
$W$ workflow hooks: Python module exposing at most the two hook points
\texttt{\small before\_query(agent, messages)} and \texttt{\small after\_observation(agent, message, outputs)};
must import and execute without error. \\
$M$ memory: Bounded list of cross-instance lessons: at most $12$ entries,
each at most $400$ characters, and any entry containing an instance-specific token (task
identifier, container or repository path) is rejected.\\
$T$ tools: Declarative shell-tool records (name, description,
in-context example, body). Each record is validated independently; invalid records are dropped and
the remainder kept. Tools are upserted by name and rendered into the system prompt as a
\texttt{\#\# Tools} section.

\paragraph{Why memory is bounded.} Memory and prompt text both reach the policy through context, so the distinction between them is not one of mechanism but of \emph{constraint}. The cap
is the mechanism: once it binds, a new lesson can be added only by merging or pruning existing
ones, which forces distillation rather than accumulation, whereas an unbounded prompt can always
absorb one more sentence. The instance-token filter enforces the other half of the contract:
memory is replayed verbatim on unseen tasks, so it must not carry instance-specific content. These
two constraints are what make the memory slot the portable component isolated in Sec.~\ref{sec:exp:universal}.

\subsection{Typed Attribution Schema: Signal Definitions}
\label{app:schema}

Each rollout receives two attributions. The \emph{deterministic bucket} is computed from
artifacts alone---no model is involved---and is the categorical signal the router and our analyses
use. The \emph{issue type} is a soft tag emitted by the analyzer for aggregation only; the
analyzer's authoritative output is a free-form root cause with artifact-anchored evidence, and it
is instructed to prefer \texttt{undetermined} over a fabricated diagnosis and \texttt{novel} when
the narrative is clear but no existing category fits.

The deterministic bucket is assigned by the first matching rule in the following fixed priority
order; the order encodes which failure \emph{blocks} the others from being observable.

\begin{enumerate}[leftmargin=*,itemsep=1pt]
\item \texttt{model\_incompat} --- the rollout failed at the API layer (message ordering or fields
the endpoint rejects, context overflow). This blocks the rollout entirely, so it outranks
everything else.
\item \texttt{budget\_exceeded} --- the rollout terminated at the step, cost, or time ceiling. An
observable budget failure, not an incorrect answer.
\item \texttt{patch\_broke\_tests} --- the grader's fix stage captured \emph{zero} tests
\emph{and} the submitted patch touches a test path. Grading applies the held-out gold test patch on
top of the agent's diff, so editing the same test file makes that application fail and the target
test never runs. This is invisible in the pass/fail boolean and mimics a clean incorrect fix, which
is why it outranks the heuristics below.
\item \texttt{unwinnable} --- the reference patch is very large (many files or bytes), i.e.\ the
instance is out of scope for a single rollout.
\item \texttt{patch\_pollution} --- the scored patch carries build artifacts, scratch files, or
frozen snapshots: a harness-pipeline self-injury rather than an agent error.
\item \texttt{no\_build} --- the command audit shows the project was never successfully built.
\item \texttt{wrong\_fix} --- default: the patch applied, the target test ran, and the behaviour is
still wrong.
\end{enumerate}

The analyzer's soft vocabulary is \texttt{loop}, \texttt{format\_violation}, \texttt{tool\_error},
\texttt{gave\_up}, \texttt{wrong\_fix}, \texttt{crash}, \texttt{budget\_exceeded},
\texttt{patch\_pollution}, \texttt{patch\_broke\_tests}, \texttt{model\_incompat},
\texttt{undetermined}, \texttt{novel}, \texttt{other}. Alongside the bucket, each rollout carries a
\emph{verification audit} (whether the project was built, whether the target test was executed
before submission) and a budget trace (steps, model calls, cost, termination reason).

\paragraph{What the buckets actually fire on.} Table~\ref{tab:buckets} gives the distribution over
all diagnosed rollouts in the grid. Two entries are worth noting. \texttt{wrong\_fix}, the class we
deliberately do \emph{not} route, is the single largest at $38.7\%$: the loop leaves more than a
third of failures untouched by construction. And \texttt{model\_incompat} is vanishingly rare
($1.0\%$), which is why no cell ever evolved a middleware.

\begin{table}[h]
\centering\small
\begin{tabular}{lrr}
\toprule
Bucket & Rollouts & Share \\
\midrule
\texttt{wrong\_fix} & 1775 & 38.7\% \\
\texttt{no\_build}               & 1514 & 33.0\% \\
\texttt{patch\_broke\_tests}     &  606 & 13.2\% \\
\texttt{budget\_exceeded}        &  525 & 11.5\% \\
\texttt{unwinnable} &   99 &  2.2\% \\
\texttt{model\_incompat}         &   44 &  1.0\% \\
\texttt{patch\_pollution}        &    9 &  0.2\% \\
\texttt{other}                   &   10 &  0.2\% \\
\midrule
Total                            & 4582 & 100\% \\
\bottomrule
\end{tabular}
\caption{Deterministic bucket distribution over all diagnosed rollouts in the grid.}
\label{tab:buckets}
\end{table}

\subsection{Routing Table: Signal to Component}
\label{app:routing}

Routing is \emph{advisory}: the attribution reaches the evolve agent as a dossier---dominant
cause, failing stages, repository state, budget trace, and competing signals---rather than as a
verdict, and the agent owns the causal claim. Table~\ref{tab:routing} gives the map it is shown.

\begin{table}[h]
\centering\footnotesize
\setlength{\tabcolsep}{2pt}
\begin{tabular}{p{3.5cm}p{3.5cm}}
\toprule
\textbf{Dominant signal} & \textbf{Prioritised component} \\
\midrule
\texttt{budget\_exceeded} & Concise memory (greps, no re-reads, minimal edits) + context trim. Raise step/cost as last resort. \\
\texttt{format\_violation} & Format-error template. \\
\texttt{loop}, missing info & Observation template. \\
Unpromptable behavior & Agent code. \\
\texttt{wrong\_fix}, \texttt{patch\_broke\_tests}, verification & Verification/boundary discipline in template or hook. \\
Missing guidance & Instance or system template. \\
\bottomrule
\end{tabular}
\caption{Advisory routing from dominant signal to harness component.}
\label{tab:routing}
\end{table}

Two constraints accompany the table. First, a \emph{soft-prompt} rule: verification is expressed
as encouragement, never as a hard gate that blocks submission, because a gate conditioned on a
check the agent may be unable to satisfy (``do not submit until the hidden test passes'') produces
non-terminating rollouts. Second, an \emph{anti-monoculture} guard: if recent rounds edited only
the instance template without a confirmed gain, the next round is required to edit a different
component. At most two components are edited per round.


\subsection{Edit Contracts, Prediction Format, and the Non-Regression Gate}
\label{app:contracts}

\paragraph{Contract format.} Every round writes a machine-readable \emph{change manifest} with
five fields: \texttt{failure\_evidence} (the specific instances and artifacts the diagnosis rests
on), \texttt{root\_cause}, \texttt{targeted\_fix} (the concrete edit), \texttt{predicted\_impact}
(an explicit list of instances the edit \emph{should} fix, and any it may regress), and
\texttt{changed\_fields}. The prediction is the falsifiable part: it commits to task-level outcomes
before they are observed.

\paragraph{Validation.} The next round re-runs the evolution split and writes a \emph{change
evaluation} against the manifest: \texttt{predicted\_fixes\_confirmed},
\texttt{predicted\_fixes\_missed}, \texttt{anticipated\_regressions},
\texttt{unattributed\_regressions}, and a \texttt{verdict}. Validation is checkpoint-level: it
scores the round's edit set as a whole, not each line independently.

\paragraph{Selection.} Let $s_r$ be the round-$r$ score on the evolution split and $\rho_r$ the
fraction of instances solved by the incumbent $H_{\text{best}}$ that round $r$ no longer solves.
The selection score is
\[
\tilde{s}_r \;=\; s_r \;-\; \lambda\,\rho_r ,
\qquad \lambda = 2.0 ,
\]
so a round that raises the aggregate while silently breaking previously-solved instances cannot
become the incumbent. Rollback is deliberately not strict, since a single-sample dip would
otherwise discard useful work: a round within a tolerance band of $0.12$ is kept without counting
against patience; the loop reverts to the incumbent only after $2$ non-improving rounds or on a
catastrophic drop of $0.25$.

\paragraph{Non-regression gate.} The harness shipped to the held-out set is the best-scoring
checkpoint, and it is floored at the seed: if no checkpoint matches or beats $H_0$ on the evolution
split, the loop returns $H_0$. The reported gain is therefore lower-bounded by zero by
construction. This one-sided guarantee is a property of editing an inspectable artifact---a file
can be reverted at file granularity---and is not available to weight-space updates.

\subsection{Meta-Model Roles and Inputs}
\label{app:prompts}

Three meta-model roles share one underlying model per configuration (the same model that serves as
the task policy $\pi$).

\paragraph{Rollout analyzer.} Runs once per failed rollout as a small tool-using agent under a
read budget of $20$ steps, with each read capped at $9{,}000$ characters, so that the diagnosis is
grounded in retrieved artifacts rather than a single long context dump. Its tools expose the
trajectory (message index, ranged reads, regex search), the submitted patch, the raw test output,
the per-target-test evaluation summary, and---critically---the grader's own stage-level verdict
(run/test/fix passed/failed/skipped counts and the broken-test-build flag). It emits the free-form
root cause, the soft issue type, secondary patterns, and a self-reported confidence.


\paragraph{Evolve agent.} Receives the layered evidence (per-rollout diagnoses aggregated by
dominant cause), the current harness contents, the evolution history with each past edit's verdict,
and the advisory routing table. It emits the change manifest of Sec.~\ref{app:contracts}. Seeing past
verdicts is what prevents it from re-proposing an already-falsified edit.

\paragraph{Distillation (Sec.~4.5 only).} A separate pass that reads the evolved memory of all
languages and returns a single bounded list of generic lessons, subject to the same admission
constraints as any memory slot plus an explicit instruction to name no build command, test path, or
file-layout convention. The result is the universal harness evaluated in Sec.~4.5.

\subsection{A Worked Round: Evidence to Diagnosis to Contract to Validation}
\label{app:worked}

We trace the C++ $\times$ Claude Haiku 4.5 cell, whose evolution-split trajectory is
$10.0 \rightarrow 62.5 \rightarrow 72.5 \rightarrow 72.5$ and whose held-out score moves from
$12.9\%$ to $87.1\%$.

\paragraph{Round 0 --- evidence.} The seed harness solves $10\%$. The deterministic buckets place
$16$ of $18$ failures in \texttt{patch\_broke\_tests}: the grader's fix stage captured zero tests
while the run stage had produced results.

\paragraph{Diagnosis.} Reading the artifacts, the analyzer reports that in these instances the
agent's source edit matches the reference fix, but the submission \emph{also} appends cases to
existing test files under \texttt{test/} and \texttt{tests/}. Grading applies the gold test patch on
top of that diff, the application conflicts, no target test is ever executed, and an otherwise
correct fix is scored zero. Note the shape of the error: the transcript ends with the agent
believing it has verified its work.

\paragraph{Contract.} The evolve agent proposes a boundary rule in the instance template and a
memory lesson naming C++ test-path conventions, and commits to a prediction: a named list of
instances that should now pass. Changed fields: instance template and memory.

\paragraph{Validation (round 2).} Re-running the split confirms the direction---the score moves
$10.0 \rightarrow 62.5$---and the change evaluation records which predicted instances actually
flipped and which did not, with no unattributed regressions; the verdict is \emph{confirmed}. Note
that the contract is scored on its own terms: several predicted instances did \emph{not} flip, and
that is recorded rather than absorbed into the aggregate.

\paragraph{Subsequent rounds.} With test-file collisions suppressed, the dominant bucket shifts to
build and verification failures, and round~2 adds a hook that builds and runs the specific failing
test before submission ($62.5 \rightarrow 72.5$). Round~3 proposes extending the boundary rule to
approval-baseline files; it does not improve the split, and the best-checkpoint rule keeps round~2
as the shipped harness. The two mechanisms in this single cell---a boundary rule and a verification
step---are the two whose relative frequency varies by language in Sec.~4.3, which is why no single
one of them is ``the'' mechanism.

\subsection{Edit-Contract Schema and a Filled Instance}
\label{app:manifest-schema}

The evolve agent's entire output is a single JSON object; nothing else it writes is read. The
schema is fixed and is the mechanism by which an edit becomes falsifiable: three fields record the
reasoning, one field commits to observable consequences, and one records what was touched.

\begin{quote}\small
\begin{description}[leftmargin=1.4em,itemsep=2pt,topsep=2pt,parsep=0pt]
\item[\ttfamily failure\_evidence] str --- the specific failing task(s) and trace evidence targeted
\item[\ttfamily root\_cause] str --- why it failed, not just what failed
\item[\ttfamily targeted\_fix] str --- what the edit changes, and why that addresses the cause
\item[\ttfamily predicted\_impact] \{"should\_pass": [task\_ids], "at\_risk": [task\_ids]\}
\item[\ttfamily edits] object keyed by editable field; only fields actually changed
\end{description}
\end{quote}

\texttt{edits} may name any of the versioned slots: \texttt{system\_template},
\texttt{instance\_template}, \texttt{observation\_template}, \texttt{format\_error\_template},
\texttt{memory}, \texttt{hooks}, \texttt{tools}. Template fields are supplied as full replacement text rather than as patches, so an
edit cannot silently corrupt a template it did not intend to touch; the result is re-rendered and
must parse before adoption. The \texttt{memory} field is additionally constrained: the agent is
shown the current memory and must return the \emph{complete} updated list, bounded to roughly a
dozen entries, so adding a lesson past the cap forces it to merge or prune rather than append.

\texttt{predicted\_impact} is the part that makes the contract falsifiable. It names instances by
identifier, before the next round is run, in two directions: those the edit should fix and those it
may break. The following round scores the manifest against observed outcomes and writes a
\emph{change evaluation} recording \texttt{predicted\_fixes\_confirmed},
\texttt{predicted\_fixes\_missed}, \texttt{anticipated\_regressions},
\texttt{unattributed\_regressions}, and a \texttt{verdict}.

\paragraph{A real contract.} Below is the verbatim manifest from round~1 of the C++ $\times$ Haiku
cell traced in Sec.~\ref{app:worked}, abridged only in the instance lists.

\begin{quote}
\begin{lstlisting}[style=compact]
{
  "failure_evidence": "16/18 failures are patch_broke_tests: catchorg__Catch2__pr-2128, pr-2187, pr-2394, pr-2723, fmtlib__fmt__pr-1171, ... nlohmann__json__pr-545. In every case the agent's source fix matches gold exactly but also edits test files (test/, tests/, test/src/), causing git apply of the gold test_patch to fail with a conflict -> broke_test_build=True -> zero tests graded -> automatic loss.",
  "root_cause": "The agent appends its own test cases to existing test files to 'demonstrate' or 'verify' its fix. The grading harness applies a gold test_patch on top of the agent's diff via `git apply`. Because the agent already modified the same test files, git apply fails with a conflict, the test build is broken, and the fix-stage captures zero test results. The source fix is often correct -- it's never evaluated because the test file collision blocks everything.",
  "targeted_fix": "Add a prominent, specific instruction in instance_template: DO NOT modify test files (files in test/, tests/, or any *test*.cpp/cc files). The harness applies its own test patch on top of your changes; if you touch test files, a merge conflict will break the entire evaluation. Only modify source/implementation files. Also add memory lessons for durability across rollouts.",
  "predicted_impact": {
    "should_pass": ["catchorg__Catch2__pr-2128", "catchorg__Catch2__pr-2187", "catchorg__Catch2__pr-2394", "catchorg__Catch2__pr-2723", "fmtlib__fmt__pr-1171", /* ... 10 more ... */ "nlohmann__json__pr-545"],
    "at_risk": ["catchorg__Catch2__pr-1608", "fmtlib__fmt__pr-2292"]
  },
  "changed_fields": ["instance_template", "memory"]
}
\end{lstlisting}
\end{quote}

Three properties of this artifact are worth noting, because they are what the instrumentation is
for. The diagnosis is \emph{anchored}: it names sixteen instances and the exact grader signature
(\texttt{broke\_test\_build}) rather than asserting a general tendency. The prediction is
\emph{two-sided}: the agent volunteers two instances it expects to harm, which is what allows a
regression to be classified as anticipated rather than unattributed. And the causal claim is
\emph{checkable against the artifact}, not against the rollout's narration---the agent in those
rollouts believed it had verified its fix.

The following round's evaluation of this contract recorded one confirmed fix among the named
instances, four predicted fixes that did not materialise, no unattributed regressions, and a
verdict of \emph{confirmed}: the direction held (the split moved $10.0 \to 62.5$) even though most
individual predictions did not. We report this rather than a cleanly-met prediction because it is
the typical case, and because it shows the contract is scored on its own terms instead of being
absorbed into the aggregate.

\subsection{A Hook and a Prompt Delta}
\label{app:g-hook-prompt}

The coding rubric of Sec.~\ref{app:rubric} treats a memory lesson, a hook, and a sentence of the
prompt delta as equivalent codeable units. Sec.~\ref{app:g-cpp} shows the memory slot; this section
shows the other two, from the same C++ $\times$ Haiku harness, so that all three slot types can be
inspected against the rubric.

\paragraph{The verification hook (round 2).} This is the edit that moves the evolution split
$62.5 \to 72.5$ in Sec.~\ref{app:worked}. It is reproduced verbatim.

\begin{quote}
\begin{lstlisting}[style=compact,language=Python]
def before_query(agent, messages) -> str | None:
    """Nudge agent to run the failing test before submitting if it hasn't done so."""
    recent = messages[-6:] if len(messages) >= 6 else messages
    recent_text = ' '.join(...)          # flatten the last few turns to text
    about_to_submit = any(kw in recent_text.lower() for kw in [
        'patch.txt', 'complete_task_and_submit', 'git diff', 'submit',
        'i have fixed', 'the fix is complete'])
    if not about_to_submit:
        return None
    has_run_test = any(kw in recent_text.lower() for kw in [
        'ctest', 'make test', 'passed', 'failed', 'error:', 'ok]', '[ ok ]',
        'running', 'test result', 'assertions', 'catch2'])
    if not has_run_test:
        return ("Before submitting: have you run the failing test to verify your fix? "
                "Try: find the build directory, run `make -j4 2>&1 | tail -20` to "
                "compile, then `ctest -R <relevant_test_name> -V 2>&1 | tail -40` to "
                "check. If you see approval/snapshot test failures, you may also need "
                "to update baseline files.")
    return None
\end{lstlisting}
\end{quote}

The hook illustrates why this slot exists as something distinct from the prompt. The rule it
enforces---verify before submitting---is already stated in the prompt, but a prompt states it
\emph{once, at the start}, whereas the hook fires \emph{at the moment the behaviour is about to
occur} and only then. It is state-dependent in both directions: it triggers on evidence that the
agent is preparing to submit, and it suppresses itself if the transcript already shows a test run.
It is also, as the routing constraint of Sec.~\ref{app:routing} requires, a \emph{nudge} and not a
gate: it returns a string that is appended to the context, and the agent remains free to submit
anyway. Under the rubric, this single unit codes to the concept \emph{run the target test before
submitting} and carries three ecosystem markers---\texttt{make -j4}, \texttt{ctest -R},
\texttt{catch2}---so it is classified as ecosystem-specific.

\paragraph{The prompt delta.} The seed's instance template is $325$ characters; the evolved one is
$1{,}730$. The added text, abridged, is:

\begin{quote}\small
\textbf{\#\# CRITICAL RULES}\\
\textbf{DO NOT modify test files} (files under \texttt{test/}, \texttt{tests/}, or any file whose
name contains \texttt{test} or \texttt{spec}). The grading system applies its own test patch\ldots\\[2pt]
\textbf{MINIMIZE your diff scope.} Only touch files that are strictly necessary for the fix. The
fewer files you modify, the less likely you are to conflict\ldots\\[2pt]
Focus your changes on:
\begin{itemize}[leftmargin=1.2em,itemsep=0pt,topsep=1pt]
\item Source files (e.g.\ \texttt{src/}, \texttt{include/}, \texttt{lib/}) that implement the buggy behavior
\item Build files only if strictly necessary to add new source files
\item Do NOT add new test cases, even to verify your fix
\end{itemize}
\textbf{\#\# Verification}\\
Before submitting, try to \textbf{run the failing test} to confirm your fix works:
\begin{enumerate}[leftmargin=1.4em,itemsep=0pt,topsep=1pt]
\item Find the failing test name from the issue or test output
\item Build and run it: e.g.\ \texttt{cd build \&\& make -j4 2>\&1 | tail -20 \&\& ctest -R <test\_name> -V 2>\&1 | tail -30}
\item If a test still fails, investigate further before submitting
\item If the repo uses approval/snapshot tests (\texttt{*.approved.txt}, \texttt{*.snap}), check whether your behavior change requires updating those baseline files
\end{enumerate}
\end{quote}

Coded under the rubric, this delta contributes units for \emph{never modify test files},
\emph{minimal diff}, \emph{localize from the failing test}, and \emph{run the target test before
submitting}---the same abstract concepts the Go harness reaches (Sec.~\ref{app:g-go})---while its
concrete tokens (\texttt{make -j4}, \texttt{ctest -R}, \texttt{src/}, \texttt{include/},
\texttt{*.approved.txt}, \texttt{*.snap}) share nothing with Go's (\texttt{go build ./...},
\texttt{go vet ./...}, \texttt{*\_test.go}, \texttt{git add}). This is the two-level structure of
Sec.~4.4 visible inside a single slot rather than aggregated over harnesses.

Note finally that the same discipline appears in all three slots of this one harness: as a rule in
the prompt, as a lesson in memory, and as a hook that fires at submission time. The loop does not
choose between them---it installs the general statement in text and the situational trigger in
code.
\section{Experimental Setup and Measurement Instruments}
\label{app:setup}

\subsection{Benchmark Splits}
\label{app:splits}

We use the eight language-specific subsets of Multi-SWE-Bench. For each language $L$ we draw a
disjoint evolution split $D^L_{\text{evo}}$ and held-out split $D^L_{\text{test}}$,
$D^L_{\text{evo}} \cap D^L_{\text{test}} = \emptyset$. The evolution split is fixed at $20$
instances per language, which is what each round is scored on. The held-out split is fixed at $50$ instances per language

The held-out split is never used to propose, score, or select an edit; it is evaluated once, at the
end, on the promoted checkpoint. All transfer and distillation experiments
(Sec.~4.5--4.6) reuse the same $D^L_{\text{test}}$, so a transplanted harness is scored on exactly
the instances the native harness was scored on.

\subsection{Model Versions and Budgets}
\label{app:budgets}

Three rollout models span distinct capability profiles: \textbf{Claude Haiku 4.5},
\textbf{GPT-5-mini}, and \textbf{DeepSeek-V4-Flash}. Within a configuration the same underlying
model serves as the task policy $\pi$ and as the outer-loop driver $\mu$, so cross-model
comparisons reflect their joint effect.

Every rollout, in every arm, runs under identical limits: a step ceiling of $80$ agent steps and a
cost ceiling of \$$1.5$ per rollout. These are hard ceilings enforced by the runtime---a rollout
that reaches either terminates and is recorded as \texttt{budget\_exceeded}
(\S\ref{app:schema})---and they apply equally to the minimal seed, mini-SWE-agent, and the evolved
harness, so no arm can buy performance with a larger budget. The evolution loop runs $3$ rounds.

\subsection{Baseline Harness Configurations}
\label{app:baselines}

\paragraph{Minimal seed $H_0$.} A stripped ReAct loop: a two-sentence system prompt stating that
the agent can interact with a shell, an instance template carrying the task text plus the
submission protocol (\texttt{git diff > patch.txt}, then an explicit completion sentinel), one bash
tool, and empty hooks and memory files. It contains no guidance about tests, builds, verification,
or diff scope---this is deliberate, since every such rule that appears in an evolved harness must
then have been derived from observed rollouts rather than inherited.

\paragraph{mini-SWE-agent.} The manually designed scaffold, run unmodified at its released
configuration, under the same step and cost ceilings, the same containers, and the same grader as
the other two arms. It is the reference for ``what a careful human scaffold achieves on this
model'', and in Sec.~4.3 its margin over $H_0$ is used as a coarse indicator of how much of a
cell's failure mass is addressable by harness design at all.

\subsection{Execution-Defect Detectors: Definitions and Precision}
\label{app:detectors}

Two automatic detectors are used in Sec.~4.3 and \S\ref{app:decomposition}. Both are computed from
artifacts only, on the \emph{minimal-seed} rollouts of the held-out set.

\begin{itemize}[leftmargin=*,itemsep=1pt]
\item \textbf{Test-file edit.} The submitted patch touches a test path. Detection is by a
language-aware matcher covering test directories (\texttt{test/}, \texttt{tests/}, \texttt{spec/},
\texttt{\_\_tests\_\_/}), Python \texttt{test\_*.py} and \texttt{conftest.py}, Go
\texttt{*\_test.go}, C/C++/Rust \texttt{*\_test.*} and \texttt{*test*.cpp}, Java CamelCase
\texttt{FooTest.java}, and JS/TS \texttt{*.test.*} and \texttt{*.spec.*}.
\item \textbf{Build break.} The grader's own record shows the fix stage captured zero tests while
the run stage produced results---i.e.\ applying the submission prevented the test executable from
being produced at all.
\end{itemize}

\paragraph{Why we report these separately rather than as one ``pollution'' rate.} The two signals
are not interchangeable, and a single combined rate is not comparable across languages. The build break is a \emph{consequence} signal: it fires when the test binary fails to exist, which a test-file edit can cause on a compiled language but which non-compiling source causes just as often. Its precision as a proxy for test-file editing therefore varies by cell---close to $1.0$ on Haiku, where nearly every break is a test edit; between $0.00$ and $1.00$ across DeepSeek's languages, where breaks on C and C++ are dominated by non-compiling source; and low on GPT-5-mini, which breaks builds without editing tests. Most importantly, on interpreted languages the build break is \emph{structurally} unable to fire---Python's rate is $0\%$ under all three models---even though agents demonstrably do edit Python test files. We therefore use the language-agnostic test-edit matcher wherever a cross-language comparison is made, and report the build break as a separate mode rather than folding the two into one number.

These two detectors cover part of the recoverable space, not all of it. The loop also installs cell-idiosyncratic rules (repository layout conventions, invocation forms, diff-scope constraints) that no detector of ours counts. Claims built on the detectors are accordingly directional: they establish that detector-covered defects account for a large share of what is recovered (\S\ref{app:decomposition}), not that they exhaust
it.

\subsection{Concept and Ecosystem-Marker Coding Rubric}
\label{app:rubric}

For the content analysis of Sec.~4.4, each evolved harness is parsed into \emph{codeable units}: a
memory lesson, a hook, or a semantically self-contained sentence of the prompt delta relative to
$H_0$. Each unit receives two independent tags.

\begin{itemize}[leftmargin=*,itemsep=1pt]
\item \textbf{Abstract concept} --- what the unit tries to enforce, independent of language:
for example \emph{do not modify test files}, \emph{localize from the failing test rather than the
issue prose}, \emph{keep the diff minimal}, \emph{verify the change builds}, \emph{run the target
test before submitting}, \emph{stage newly created files}.
\item \textbf{Ecosystem marker} --- any concrete token naming the target's machinery: a build or
test command (\texttt{make -j4}, \texttt{ctest -R}, \texttt{go build ./...}, \texttt{mvn test -Dtest=}),
a test-path glob (\texttt{*\_test.go}, \texttt{*test*.cpp}), a language-specific file or layout
convention (\texttt{single\_include/}, \texttt{*.approved.txt}, \texttt{conftest.py}), or an
idiom tied to the language (\texttt{errors.Is}).
\end{itemize}

A unit is \emph{ecosystem-specific} if it names at least one marker. Cross-language overlap is then
measured at each level as the Jaccard similarity between the two languages' tag sets, and the
ecosystem-specific fraction of a harness is the share of its units carrying a marker.

Two properties of this instrument bound what it can show. Concept tags come from an LLM rubric with
imperfect granularity, which splits near-synonymous concepts and therefore \emph{deflates} concept
overlap; and marker extraction is lexical, so a discipline expressed without naming any concrete
token is counted as generic even if it was learned from one ecosystem. Both biases push the two
levels toward each other, so the \emph{gap} between concept overlap and marker overlap---which is
what Sec.~4.4 rests on---is if anything understated. We do not interpret the absolute value at
either level.

\section{Full Results}
\label{app:results}
\subsection{Held-Out \texttt{mean\_solve@3} per Cell}
\label{app:percell}

Table~\ref{tab:percell} gives every number behind Fig.~\ref{fig:main}: the minimal seed $H_0$,
mini-SWE-agent, and the evolved harness, for all $8\times3$ cells. All values are held-out
$\mathrm{mean\_solve@3}$.

\begin{table}[ht]
\centering\small
\setlength{\tabcolsep}{4pt}
\begin{tabular}{llccc c}
\toprule
Model & Lang & $H_0$ & mini-SWE & TRIAGE & $\Delta$ vs $H_0$ \\
\midrule
\multirow{8}{*}{Haiku 4.5}
 & C++    & 12.9 & 61.9 & \textbf{87.1} & $+74.3$ \\
 & Java   &  2.9 & 51.4 & \textbf{82.9} & $+80.0$ \\
 & C      & 30.0 & 75.0 & \textbf{81.7} & $+51.7$ \\
 & JS     & 31.5 & 68.5 & \textbf{75.9} & $+44.4$ \\
 & TS     & 34.6 & 65.4 & \textbf{86.5} & $+51.9$ \\
 & Rust   & 11.1 & 31.5 & \textbf{72.2} & $+61.1$ \\
 & Go     & 13.0 & 25.9 & \textbf{37.0} & $+24.1$ \\
 & Python & 75.0 & 72.6 & 73.2 & $-1.8$ \\
\midrule
\multirow{8}{*}{GPT-5-mini}
 & C++    & 22.9 & \textbf{27.6} & 27.1 & $+4.3$ \\
 & Java   & 22.9 & 16.2 & \textbf{28.6} & $+5.7$ \\
 & C      & 28.3 & \textbf{46.7} & 36.7 & $+8.3$ \\
 & JS     & 33.3 & \textbf{37.0} & 31.5 & $-1.8$ \\
 & TS     & 46.2 & 36.5 & 44.2 & $-2.0$ \\
 & Rust   & 24.1 & \textbf{31.5} & 20.4 & $-3.7$ \\
 & Go     & 22.2 & \textbf{27.8} & 22.2 & $\pm0.0$ \\
 & Python & 67.9 & 65.5 & 67.9 & $\pm0.0$ \\
\midrule
\multirow{8}{*}{DS-V4-Flash}
 & C++    & 34.3 & 60.0 & \textbf{65.7} & $+31.4$ \\
 & Java   & 24.3 & 35.7 & \textbf{54.3} & $+30.0$ \\
 & C      & 38.3 & 65.0 & 65.0 & $+26.7$ \\
 & JS     & 38.9 & 46.3 & \textbf{59.3} & $+20.4$ \\
 & TS     & 44.2 & \textbf{65.4} & 63.5 & $+19.3$ \\
 & Rust   & 18.5 & \textbf{44.4} & 35.2 & $+16.7$ \\
 & Go     & 18.5 & 27.8 & \textbf{33.3} & $+14.8$ \\
 & Python & 73.2 & 71.4 & 66.1 & $-7.1$ \\
\bottomrule
\end{tabular}
\caption{Held-out $\mathrm{mean\_solve}$ (\%) per cell. Bold marks the best of the three arms.}
\label{tab:percell}
\end{table}

\subsection{Stability and Significance}
\label{app:stability}

Table~\ref{tab:sig} reports, per cell, the aggregate $\pm$ its standard deviation across resamples
($\sigma = s/\sqrt{n}$), the paired improvement, and the paired statistic
$z = \Delta/\sigma_\Delta$. Stars denote $|z|\ge 1.64$ ($^{*}$), $\ge 1.96$ ($^{**}$),
$\ge 2.58$ ($^{***}$).

The pattern is the one the mechanism predicts. Every non-Python cell is significant for the two
models with substantial defect mass (Haiku all $^{***}$, $z=3.3$--$11.7$; DeepSeek $z=1.7$--$4.6$),
and \emph{no} GPT-5-mini cell reaches significance ($|z|<1.3$)---which is why we read that region
as a null rather than as a gain. Python is null or slightly negative under all three models.

\begin{table}[ht]
\centering\small
\setlength{\tabcolsep}{4pt}
\begin{tabular}{lcccc}
\toprule
Lang  & $H_0$ & TRIAGE & $\Delta$ (pp) & $z$ \\
\midrule
\multicolumn{5}{l}{\emph{Claude Haiku 4.5}} \\
\quad C++  & 12.9\,$\pm$\,5.2 & 87.1\,$\pm$\,4.7 & $+74.3$\,$\pm$\,7.2 & $+10.30^{***}$ \\
\quad Java & 2.9\,$\pm$\,2.0 & 82.9\,$\pm$\,6.1 & $+80.0$\,$\pm$\,6.9 & $+11.66^{***}$ \\
\quad C & 30.0\,$\pm$\,7.8 & 81.7\,$\pm$\,6.6 & $+51.7$\,$\pm$\,8.8 & $+5.87^{***}$ \\
\quad Go & 13.0\,$\pm$\,6.3 & 37.0\,$\pm$\,8.7 & $+24.1$\,$\pm$\,7.2 & $+3.32^{***}$ \\
\quad Rust & 11.1\,$\pm$\,6.2 & 72.2\,$\pm$\,8.2 & $+61.1$\,$\pm$\,9.0 & $+6.80^{***}$ \\
\quad TS & 34.6\,$\pm$\,8.7 & 86.5\,$\pm$\,4.4 & $+51.9$\,$\pm$\,8.5 & $+6.08^{***}$ \\
\quad JS & 31.5\,$\pm$\,8.1 & 75.9\,$\pm$\,7.2 & $+44.4$\,$\pm$\,8.6 & $+5.18^{***}$ \\
\quad Python & 75.0\,$\pm$\,7.5 & 73.2\,$\pm$\,7.9 & $-1.8$\,$\pm$\,6.5 & $-0.27$ \\
\midrule
\multicolumn{5}{l}{\emph{GPT-5-mini}} \\
\quad C++  & 22.9\,$\pm$\,6.3 & 27.1\,$\pm$\,6.9 & $+4.3$\,$\pm$\,4.8 & $+0.90$ \\
\quad Java & 22.9\,$\pm$\,6.3 & 28.6\,$\pm$\,6.6 & $+5.7$\,$\pm$\,4.9 & $+1.16$ \\
\quad C & 28.3\,$\pm$\,6.6 & 36.7\,$\pm$\,7.6 & $+8.3$\,$\pm$\,6.8 & $+1.22$ \\
\quad Go & 22.2\,$\pm$\,7.7 & 22.2\,$\pm$\,7.7 & $\pm0.0$\,$\pm$\,4.6 & $0.00$ \\
\quad Rust & 24.1\,$\pm$\,7.2 & 20.4\,$\pm$\,6.1 & $-3.7$\,$\pm$\,7.5 & $-0.49$ \\
\quad TS  & 46.2\,$\pm$\,9.2 & 44.2\,$\pm$\,8.5 & $-1.9$\,$\pm$\,7.1 & $-0.27$ \\
\quad JS  & 33.3\,$\pm$\,9.2 & 31.5\,$\pm$\,8.5 & $-1.9$\,$\pm$\,5.0 & $-0.37$ \\
\quad Python  & 67.9\,$\pm$\,8.2 & 67.9\,$\pm$\,8.2 & $\pm0.0$\,$\pm$\,3.6 & $0.00$ \\
\midrule
\multicolumn{5}{l}{\emph{DeepSeek-V4-Flash}} \\
\quad C++ & 34.3\,$\pm$\,6.7 & 65.7\,$\pm$\,7.0 & $+31.4$\,$\pm$\,6.8 & $+4.61^{***}$ \\
\quad Java & 24.3\,$\pm$\,6.9 & 54.3\,$\pm$\,7.5 & $+30.0$\,$\pm$\,6.9 & $+4.37^{***}$ \\
\quad C  & 38.3\,$\pm$\,7.5 & 65.0\,$\pm$\,7.6 & $+26.7$\,$\pm$\,7.5 & $+3.57^{***}$ \\
\quad Go & 18.5\,$\pm$\,7.1 & 33.3\,$\pm$\,9.2 & $+14.8$\,$\pm$\,7.5 & $+1.99^{**}$ \\
\quad Rust & 18.5\,$\pm$\,7.1 & 35.2\,$\pm$\,7.9 & $+16.7$\,$\pm$\,9.6 & $+1.73^{*}$ \\
\quad TS & 44.2\,$\pm$\,8.9 & 63.5\,$\pm$\,8.1 & $+19.2$\,$\pm$\,8.8 & $+2.18^{**}$ \\
\quad JS & 38.9\,$\pm$\,9.0 & 59.3\,$\pm$\,9.3 & $+20.4$\,$\pm$\,8.1 & $+2.51^{**}$ \\
\quad Python & 73.2\,$\pm$\,7.9 & 66.1\,$\pm$\,8.2 & $-7.1$\,$\pm$\,4.2 & $-1.69^{*}$ \\
\bottomrule
\end{tabular}
\caption{Per-cell stability and paired significance. Values are mean $\pm$ resample standard
deviation; $\Delta$ is the paired improvement over $H_0$ with its own $\sigma_\Delta$.}
\label{tab:sig}
\end{table}

\subsection{Cross-Language Jaccard: DeepSeek-V4-Flash}
\label{app:jaccard-ds}

\begin{figure}[ht]\centering
\includegraphics[width=\linewidth]{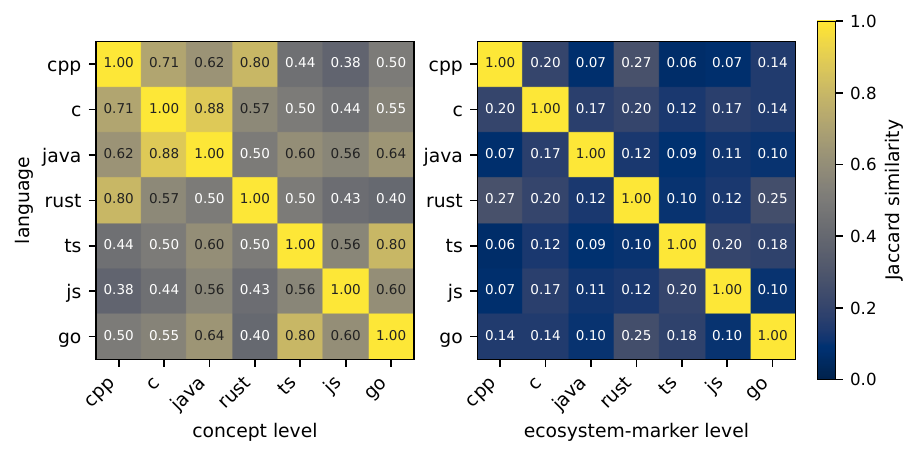}
\caption{Cross-language harness overlap for DeepSeek-V4-Flash. Left: abstract concepts (mean
off-diagonal $0.57$). Right: concrete ecosystem markers (mean off-diagonal $0.14$).}
\label{fig:jaccard-ds}
\end{figure}

Fig.~\ref{fig:jaccard-ds} repeats the content analysis of Sec.~4.4 on the second model with full
language coverage. The two-level structure replicates: mean off-diagonal concept overlap $0.57$
(Haiku: $0.55$) against ecosystem-marker overlap $0.14$ (Haiku: $0.12$), a factor of four. That the
gap reproduces under a different rollout model is the reason we read it as a property of the
language axis rather than of one model's writing style.

\subsection{Ecosystem-Specific Fraction: Per-Cell Breakdown}
\label{app:ecofrac}

Table~\ref{tab:ecofrac} gives the per-cell values summarised in Fig.~5. Every evolved harness
falls in the $20$--$40\%$ band, with cell means of $26\%$ (Haiku) and $29\%$ (DeepSeek). The
ordering across languages is not identical between the two models---C is the most
ecosystem-specific cell for DeepSeek ($40\%$) but mid-range for Haiku ($25\%$)---so we read the
\emph{band}, not the per-language ranking, as the stable quantity.

\begin{table}[ht]
\centering\small
\begin{tabular}{lccccccc|c}
\toprule
Model & C++ & Java & C & Go & Rust & TS & JS & mean \\
\midrule
Haiku        & 29 & 29 & 25 & 24 & 31 & 27 & 18 & 26 \\
DS & 24 & 25 & 40 & 30 & 33 & 30 & 20 & 29 \\
\bottomrule
\end{tabular}
\caption{Ecosystem-specific fraction (\%) of each evolved harness.}
\label{tab:ecofrac}
\end{table}

\section{Mechanism Analyses}
\label{app:mechanism}

\subsection{Failure-Type Decomposition of Recovered Instances}
\label{app:decomposition}

Sec.~4.3 claims that the gains compensate recoverable execution defects. The most direct test is to
take the instances the evolved harness actually \emph{recovered}---those it solves in a strictly
larger fraction of its $k$ rollouts than the seed did---and ask what the seed's failure looked like
on exactly those instances. We classify each recovered instance by the seed's behaviour using the
two detectors of Sec.~\ref{app:detectors}.

Across the grid, $256$ held-out instances are recovered. Of these, $88\%$ carried a
detector-covered defect under the seed: $81\%$ had a submitted patch touching a test path, and a
further $7\%$ broke the test build without touching a test file. The remaining $12\%$ show neither
signature and are recoveries we cannot attribute to a detected defect.

\begin{table}[ht]
\centering\small
\begin{tabular}{lcccc}
\toprule
Model & recovered & test-edit & build-break & neither \\
\midrule
Haiku 4.5         & 142 & 94\% &  1\% &  4\% \\
DS-V4-Flash &  81 & 81\% & 10\% &  9\% \\
GPT-5-mini        &  33 & 21\% & 27\% & 52\% \\
\midrule
Pooled            & 256 & 81\% &  7\% & 12\% \\
\bottomrule
\end{tabular}
\caption{What the minimal seed was doing on the instances the evolved harness recovered.}
\label{tab:decomposition}
\end{table}

\begin{figure}[ht]\centering
\includegraphics[width=\linewidth]{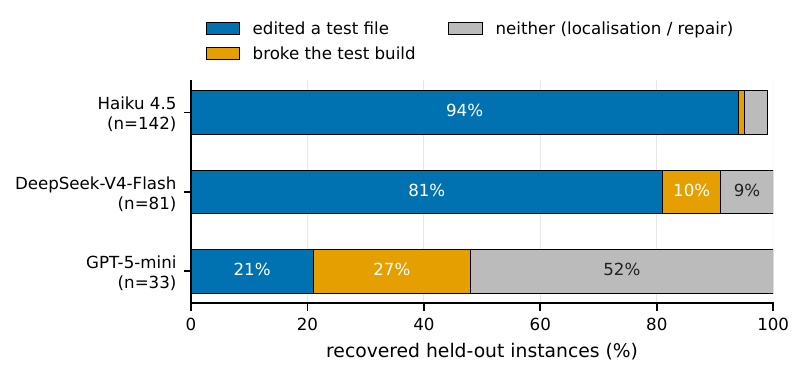}
\caption{Seed-side failure signature of recovered held-out instances, by model.}
\label{fig:decomposition}
\end{figure}

Two readings follow. For the two models with substantial gains, recovery is overwhelmingly the
suppression of a self-inflicted defect: $95\%$ (Haiku) and $91\%$ (DeepSeek) of recovered instances
were failing for a reason the harness can state as a rule. For GPT-5-mini the picture inverts---a
majority ($52\%$) of its few recoveries show neither signature---which is the same fact reported in
Sec.~4.3 from the other direction: with little detector-covered defect mass to remove, what
little it gains is not compensation of the kind we can attribute.

\paragraph{Relation to the ceiling claim.} This decomposition is about \emph{recovered} instances
and therefore says nothing about instances that remained unsolved. It supports the directional
claim (compensating recoverable execution defects is a principal source of the gains) and not the
stronger one we explicitly decline in Sec.~4.3 (that detected defect mass bounds the gain available
in a cell).

\subsection{Gain Against Defect Mass}
\label{app:defectmass}

\begin{figure}[ht]\centering
\includegraphics[width=0.8\linewidth]{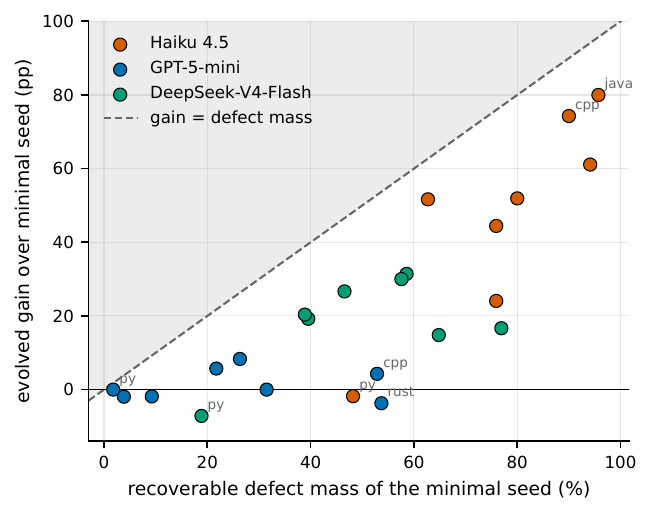}
\caption{Seed defect mass against evolved gain, one point per $\langle L,\pi\rangle$ cell,
coloured by model. The diagonal marks equality; the shaded region is where no cell falls.}
\label{fig:defectgain}
\end{figure}

Fig.~\ref{fig:defectgain} plots, for each of the $24$ cells, the seed's detector-covered defect
rate against the net gain of the evolved harness. The association is strong
($r = 0.81$), and it is the relationship the compensation account predicts: cells whose base
policy sabotages itself often are the cells where installing a rule pays.

Three qualifications matter, and we state them rather than smoothing them over.
(i) \emph{The relationship is not a bound we claim.} No cell converts more of its defect mass than
it has, but our detectors cover only part of the recoverable space (Sec.~\ref{app:detectors}), so the
$x$-axis is a measured indicator and not an estimate of headroom.
(ii) \emph{Conversion is not constant.} The gain per unit of defect mass differs by model---about
$0.58$ for Haiku, $0.34$ for DeepSeek, $0.12$ for GPT-5-mini---so defect mass alone does not
determine the gain; how much of it a policy can be induced to give up is a second factor.
(iii) \emph{Within-model association varies.} It is strong for Haiku ($r=0.85$) and moderate for
DeepSeek ($r=0.54$), but absent for GPT-5-mini ($r=0.12$), where the gain range is too compressed
($-3.7$ to $+8.3$~pp) for any relationship to be visible. The pooled $r=0.81$ is therefore carried
by the two models that have gains at all.

\subsection{What the Loop Encodes: A Planted-Defect Probe}
\label{app:planted}

The analyses above are observational: they describe what the loop wrote when left to find its own
targets. This probe asks a sharper question---can the loop discover and persist an operational fact
that is \emph{not} already part of general good practice?

We construct a controlled defect that is observable and non-shortcut-able. A gate is injected into
the environment such that the obvious action (running the failing test) fails with an explicit
error naming the remedy, and the same gate is applied by the grader, so the failure is real rather
than cosmetic. Crucially, no amount of generic discipline helps: localizing from the traceback,
minimising the diff, or running the test are all \emph{blocked} by the gate rather than assisted by
it. The only remedy is the specific operational fact, which surfaces only after running once and
reading the error, and which persists through a new non-test file so that no test-pollution
collision confounds the measurement.

The result is negative. Across the probe conditions, the loop did not converge on the planted fact:
held-out scores moved $78.3\% \to 76.1\%$ in the control and $76.1\% \to 71.7\%$ when the
environment additionally surfaced the error text, both within noise and both non-positive. Instead
of encoding the specific operational fact, the loop wrote generic verification discipline of the
kind it writes everywhere else.

We report this because it bounds the claim of the paper in a useful direction. The loop is
effective at \emph{recall}---recognising a recurring failure mode and installing the discipline that
suppresses it---and we have no evidence that it performs \emph{discovery} of novel, environment-
specific operational knowledge. That is consistent with everything else we measure: the content is
a playbook of general disciplines instantiated on a concrete ecosystem (Sec.~4.4), the portable
part distills cleanly (Sec.~4.5), and what does not transfer is ecosystem machinery rather than
insight.

\section{Ablations}
\subsection{Disentangling Meta-Agent Capability from Harness Gains}
\label{sec:meta_agent_ablation}

A potential concern is whether the performance gains come from external domain knowledge or superior reasoning injected by the strong Meta-Agent rather than the self-evolution mechanism itself. To strictly isolate this factor, we conduct a controlled ablation on a representative, medium-complexity cell: \texttt{(Go, DeepSeek-V4-Flash)}. Holding the rollout base policy (\texttt{DeepSeek-V4-Flash}) fixed, we evaluate the harness evolution loop under three distinct Meta-Agent drivers spanning a wide spectrum of model capabilities:
\begin{enumerate}[leftmargin=*]
    \item \textbf{Strong Meta-Agent}: Claude Sonnet 4.6 (a strong model).
    \item \textbf{Medium-Weak Meta-Agent}: \texttt{DeepSeek-V4-Flash} (a native, same-model self-evolution loop).
    \item \textbf{Ultra-Weak Meta-Agent}: \texttt{GPT-OSS-20B} (an open-weights 20B-parameter model).
\end{enumerate}

\begin{table}[ht]
\centering
\small
\begin{tabular}{lccc}
\toprule
\textbf{Meta-Agent Driver} & \textbf{Held-Out}  \\
\midrule
Minimal Seed $H_0$  & 18.5  \\
\midrule
\texttt{GPT-OSS-20B} & 34.2 \\
\texttt{DeepSeek-V4-Flash}  & 33.3 \\
Claude Sonnet 4.6  & 35.6  \\
\bottomrule
\end{tabular}
\caption{Meta-Agent driver ablation on the \texttt{(Go, DeepSeek-V4-Flash)} cell ($k=3$ rollouts). Even an ultra-weak 20B open Meta-Agent recovers the performance gains delivered by the frontier model, confirming that harness evolution is governed by structural diagnostic routing rather than external knowledge injection.}
\label{tab:meta_agent_ablation}
\end{table}

As presented in Table~\ref{tab:meta_agent_ablation}, all three Meta-Agents consistently suppress self-inflicted test pollution. Qualitative inspection of the evolved artifacts confirms that even the 20B model successfully extracts the correct behavioral contracts. These results provide evidence that the primary driver of capability recovery is \textbf{not the extra external knowledge or advanced reasoning of a superior Meta-Agent}, but rather the \textit{diagnostic-routed feedback loop} and \textit{falsifiable contract system}. By converting raw execution traces into explicit, typed failure signals (Sec.~\ref{sec:method}), the task complexity of harness editing is drastically simplified—allowing even an open 20B model to reliably identify failure modes and write effective behavioral constraints.

To be clear, we do not claim that the Meta-Agent's capability is entirely irrelevant. A baseline reasoning threshold is inherently required for the Meta-Agent to follow instruction schemas, digest structural telemetry, and write syntactically sound harness edits. However, our findings demonstrate that once the Meta-Agent satisfies this minimal operational threshold (e.g., at the level of standard open 20B models), the structural design of the evolution framework—namely, diagnostic failure routing and falsifiable contract enforcement—becomes the primary governing factor.

\subsection{Capability Tools Are Reachable Behaviourally}
\label{app:ablations}
\label{app:ftdup}

We observe that the loop essentially never writes tools. One reading is that
tools are unnecessary; another is that our substrate makes them hard to reach. We test the
distinction by supplying the tool ourselves.

We take a task family where a custom capability is plausibly useful---detecting duplicated code
regions during repair---implement \texttt{find\_duplicates} as a first-class tool, and pre-install
it in the harness so the loop begins with the capability rather than having to invent it. We then
evolve with and without the tool present, holding everything else fixed.

\begin{table}[ht]
\centering\small
\begin{tabular}{lccc}
\toprule
Arm & seed & evolved & $\Delta$ \\
\midrule
Haiku, no tool          & 11.1 & \textbf{22.2} & $+11.1$ \\
Haiku, tool pre-installed & 11.1 & 9.3 & $-1.8$ \\
\bottomrule
\end{tabular}
\caption{Held-out $\mathrm{mean\_solve}$ (\%) with and without a pre-installed capability tool.}
\label{tab:ftdup}
\end{table}

The tool is net-negative: supplying the capability up front does not merely fail to
help, it costs $12.9$~pp (Haiku \& C++) relative to evolving without it. The
mechanism is visible in the traces---the tool competes for steps and attention with the behavioural
disciplines that actually carry the gain, and the loop spends rounds integrating it instead of
installing rules. We draw a narrow conclusion: within this substrate and budget, the absence of
evolved tools reflects their low marginal value against behavioural edits, not an inability of the
loop to reach them.

\section{Qualitative Evolution Trace}
\label{app:traces}

The trace below plots the evolution-split score by round, annotated with what that round changed,
the verdict the following round assigned to it, and the concrete machinery the harness named.
Together they illustrate the claim of Sec.~4.3 that which defect binds is a property of the cell:
the same loop writes a boundary rule where test pollution dominates and a verification step where
it does not.

\label{app:trace-cpp}
\begin{figure}[ht]\centering
\includegraphics[width=\linewidth]{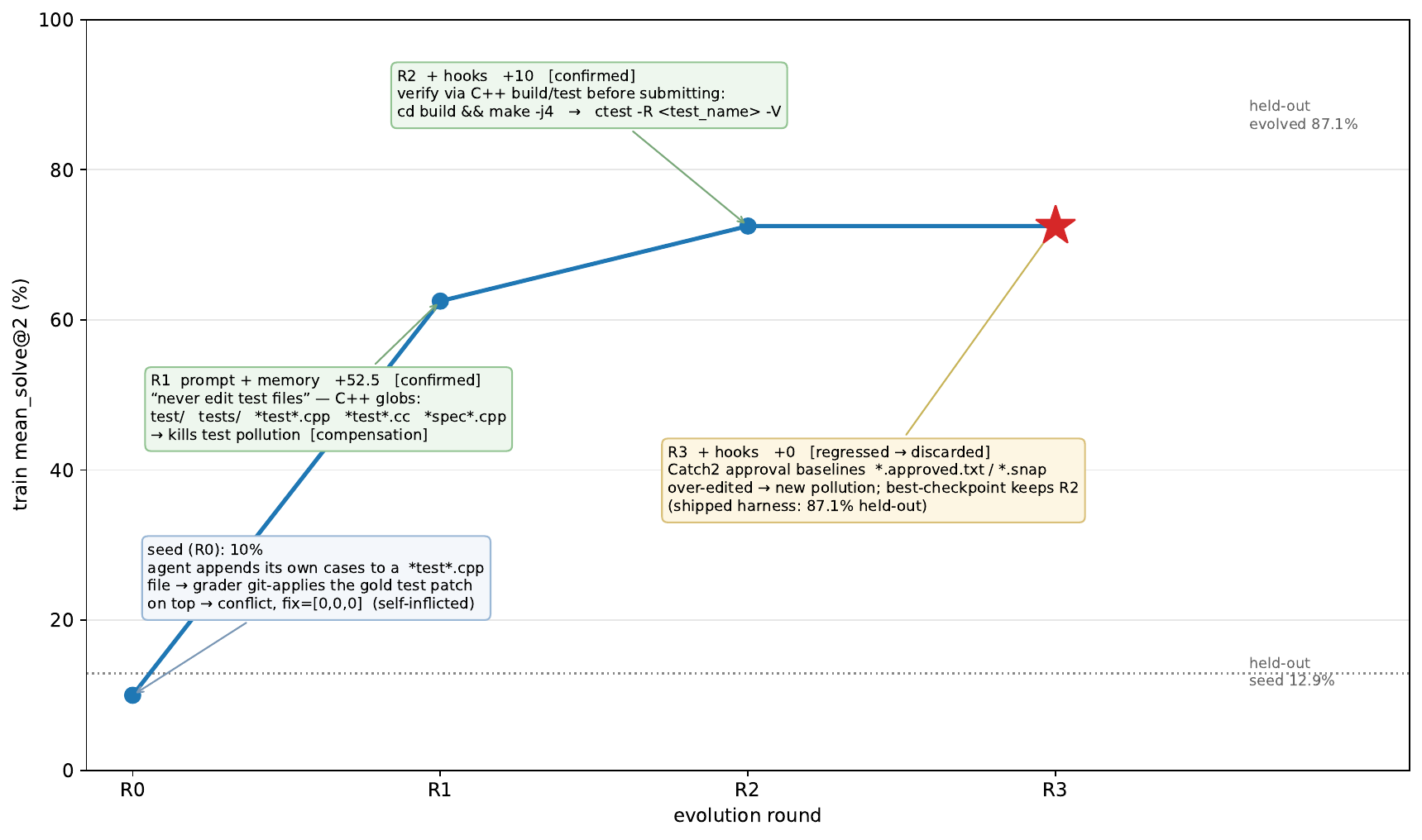}
\caption{C++ $\times$ Haiku. Round~1 installs the test-file boundary and moves the split
$10.0 \to 62.5$; round~2 adds build-and-run-the-target-test verification; round~3 over-extends the
boundary to approval baselines, does not improve, and is discarded by the best-checkpoint rule.
Held-out: $12.9 \to 87.1$.}
\label{fig:trace-cpp}
\end{figure}

\section{Example Evolved Harnesses}
\label{app:examples}

We reproduce two evolved harnesses in full, chosen to make the Sec.~4.4 claim inspectable: they
come from the \emph{same} rollout model and differ only in language, so any divergence is
attributable to the ecosystem. Both are dominated by the same abstract concepts---do not touch
tests, minimise the diff, localize from the failing test, verify before submitting---while sharing
almost no concrete token.

\subsection{C++ plus Haiku (memory slot, full text)}
\label{app:g-cpp}
\begin{quote}\small\ttfamily
\begin{itemize}[leftmargin=1em,itemsep=1pt,label={--}]
\item NEVER modify test files (test/, tests/, *test*.cpp, *test*.cc, *spec*.cpp). The grading
harness applies its own test patch on top of your diff --- editing test files causes a merge
conflict, breaks the test build, and scores your correct source fix as zero.
\item MINIMIZE your diff scope: only touch files strictly necessary for the fix. The grading
harness also modifies some source files via its test patch --- touching extra files increases the
risk of a merge conflict.
\item Run the failing test before submitting. Build first (e.g.\ \texttt{make -j4}), then run the
specific failing test (e.g.\ \texttt{ctest -R <test\_name> -V}).
\item Localize from the failing test or traceback first; do not guess from issue prose alone.
\item When a repo has approval/snapshot baseline files (\texttt{*.approved.txt}, \texttt{*.snap},
\texttt{*.baseline}), changing behavior may require updating those baselines too.
\item When fixing a bug that involves control flow, trace the full call graph to understand when
each function runs; an existing call may need to be removed before adding new ones.
\item When a PR adds a new feature, cover the full scope of the PR, not just one aspect.
\item After making a fix, verify it compiles: \texttt{cd <build\_dir> \&\& make -j4}.
\item If conditional output behavior is involved (e.g.\ an \texttt{ensure\_ascii} flag), handle all
relevant branches, not just the default case.
\end{itemize}
\end{quote}

\subsection{Go $\times$ Haiku (memory slot, full text)}
\label{app:g-go}
\begin{quote}\small\ttfamily
\begin{itemize}[leftmargin=1em,itemsep=1pt,label={--}]
\item NEVER modify \texttt{*\_test.go} files or test fixture files (\texttt{test/fixtures/*}). The
grading harness applies its own gold test patch on top of your diff --- if you edited those files,
\texttt{git apply} will conflict and your fix will never be graded.
\item If your fix requires creating NEW files, run \texttt{git add <newfile>} before generating the
patch; plain \texttt{git diff} only shows already-tracked files.
\item Localize from the failing test name and error trace, not just the issue prose.
\item Make the minimal correct change; avoid refactoring surrounding code unnecessarily.
\item Use targeted reads: grep for the relevant symbol first, then read only the pertinent lines.
Avoid \texttt{cat}-ing whole large files.
\item Prefer \texttt{errors.Is()} over \texttt{!=} for error comparisons in Go; use the exact
output format strings expected by existing tests.
\item When adding a feature, follow an existing parallel function/pattern exactly --- naming,
signature, error-handling style.
\item After making a change, verify it compiles: \texttt{go build ./...} or \texttt{go vet ./...}.
\item Submit with \texttt{git add -A \&\& git diff HEAD > patch.txt}, then the completion sentinel.
This captures both modified and new files.
\end{itemize}
\end{quote}

Read side by side, the two harnesses make the two-level structure concrete. The abstract concepts
recur almost one for one. The instantiations share nothing: \texttt{make -j4} and \texttt{ctest -R}
against \texttt{go build ./...} and \texttt{go vet ./...}; \texttt{*test*.cpp} against
\texttt{*\_test.go}; C++ approval baselines against Go's untracked-file staging problem; and one
language-idiom rule (\texttt{errors.Is}) with no C++ counterpart. Note also that the Go harness
spends a lesson on \emph{token efficiency} (targeted reads) that C++ does not---Go's cell hits the
budget ceiling more often---which is the efficiency channel of Sec.~\ref{app:schema} appearing in the
artifact.

\subsection{The Distilled Universal Harness (ten guidelines)}
\label{app:g-universal}

This is the artifact evaluated in Sec.~4.5. It is memory-only and, by construction, names no build
command, test path, or file-layout convention.

\begin{quote}\small
\begin{enumerate}[leftmargin=1.4em,itemsep=1pt]
\item Never modify test files; the grading harness applies its own test patch on top of your diff,
and any conflict causes a correct source fix to score zero.
\item Make the minimal diff: touch only the source files strictly necessary for the fix.
\item Localize from the failing test output and stack trace first, not from the issue prose alone.
\item Before submitting, verify your fix compiles or type-checks by running the language's build
step; an uncompiled fix is unverified.
\item Run the specific failing test after applying your fix to confirm it passes before submitting.
\item If you create new files, explicitly stage them before generating your patch; untracked files
are invisible to a plain diff.
\item Never rename or move existing files; the grader's patch targets original paths.
\item When an issue covers multiple behaviors, implement all required changes---partial fixes fail
the remaining cases.
\item When fixing control-flow or state-management bugs, trace the full call graph to understand
ordering; adding a call may require removing or guarding an existing one.
\item Before finalizing, re-read the failing test assertions to confirm the implementation matches
exactly what the test expects.
\end{enumerate}
\end{quote}

Comparing this list with the two harnesses above shows what distillation keeps and what it drops.
The concepts survive nearly intact. What is removed is every concrete token---and that removal is
what Fig.~\ref{fig:universal} prices: on generic targets it costs nothing, and on ecosystem-heavy ones it costs
$32$--$52\%$ of the native gain.

\section{Limitations}
\label{sec:limitations}

\paragraph{Task scope, and the parts of the design it leaves untested.}
Our evidence comes entirely from Multi-SWE-Bench-style tasks: one well-specified repair goal
per instance, resolved in a single episode and verified by a fixed test suite. This regime
does not exercise multi-goal or long-horizon behaviour, and two components of the evolvable
space are correspondingly under-exercised. Persistent memory is tested only as distilled
cross-instance discipline, never as within-episode state carried across many subgoals;
sub-agent decomposition has little to do in tasks whose plan is typically a handful of
steps. The loop is not specialized to this format---nothing in diagnostic routing or the
contract system assumes single-goal tasks---but we do not demonstrate that it transfers to
longer horizons, and we expect the balance between behavioural and structural edits to shift
if it does.

\paragraph{Dependence on the grading protocol.}
What counts as a \emph{recoverable defect} (Sec.~\ref{sec:exp:mechanism}) is partly defined
by how the benchmark scores. Editing a test file is fatal here because the grader applies a
hidden gold test patch on top of the agent's diff; in a deployment without that convention
the same edit would be merely questionable, not automatically zero. Two of the three
measured defects---shipping non-building source, submitting without running the target
test---are protocol-independent engineering failures, but the magnitude of the first is
specific to SWE-Bench-style grading. We therefore read the absolute gain as partly reflecting
what the harness learns about the evaluation protocol, and the \emph{comparison} across
harnesses, which holds the protocol fixed, as the reliable quantity.

\paragraph{No head-to-head comparison with related harness-optimization methods.}
We do not claim novelty for the general idea of automatically editing an agent's scaffold;
our design shares its core with concurrent work~\citep{ahe, harnessfix}, and our contribution
is a lightweight, plug-in instantiation together with the analysis of \emph{when and why}
such optimization pays off. We nonetheless do not run those methods as baselines: each
requires its own evolution budget per cell, and reproducing them across a $8\times3$ grid was
outside our compute budget. Two further comparisons are missing for the same reason: stronger
manually designed scaffolds than \texttt{mini-SWE-agent}, and a compute-matched inference-time
baseline that spends the evolution budget on additional rollouts instead. The last is the
most informative gap---we show the harness beats a fixed scaffold at equal inference cost,
not that harness evolution is the best use of a fixed \emph{total} budget.

\paragraph{Limited component-level attribution.}
We isolate one design axis---the capability of the meta-agent driving the loop
(Sec.~\ref{sec:meta_agent_ablation})---and find that gains survive an ultra-weak driver, which
rules out knowledge injection from a frontier model as the source of the improvement. We do
not ablate the remaining elements individually: diagnostic routing, the falsifiable-contract
format, the non-regression gate, and two-axis editing are evaluated only as a bundle. The
evidence therefore supports the claim that this bundle works and that its benefit is not
attributable to the meta-agent's strength, but it does not apportion credit among the
bundle's parts.

\paragraph{Statistical power and search budget.}
Cost constraints bound the protocol in four ways. Splits are small (20 evolution / 50
held-out instances per language), so per-cell
numbers are individually noisy and we interpret aggregate structure---marginals, sign
patterns, regime boundaries---rather than single values. Evolution runs once per cell, so we
report rollout variance but not variance over the loop's own stochasticity; the
cell-specificity of Sec.~\ref{sec:exp:specificity} is consequently a lower bound on
agreement, since run-to-run variation is folded into the cross-language differences. The loop
runs three rounds of greedy hill-climbing rather than a population-based search, so the
harnesses we report are reachable improvements, not optima. Finally, our defect detectors and
concept/ecosystem rubric are automated instruments with imperfect precision; we rely on them for relative composition across cells rather
than for absolute defect rates.

\section{Computational Resources and Cost}
\label{app:compute}

\subsection{Scale of the Experiments}
\label{app:compute-scale}

The unit of compute is a \emph{rollout}: one agent episode on one instance, bounded at $80$ steps
and \$$1.5$ (Sec.~\ref{app:budgets}), followed by containerised grading with the benchmark's own
harness. Table~\ref{tab:compute} accounts for every rollout executed for this paper, recovered from
the per-rollout records rather than estimated.

\begin{table*}[ht]
\centering\small
\setlength{\tabcolsep}{5pt}
\begin{tabular}{lrrr}
\toprule
Component & Rollouts & Agent steps & Metered cost \\
\midrule
Main grid, $8\times3$ cells (\S\ref{app:compute-scale}) &  7{,}053 & 330{,}060 & \$1{,}415 \\
\quad --- evolution loop                     &  4{,}233 &          &        \\
\quad --- held-out evaluation                &  2{,}820 &          &        \\
Cross-language transfer (Sec.~4.6)           &  4{,}792 & 237{,}964 & \$1{,}088 \\
mini-SWE-agent baseline                      &  1{,}606 &  72{,}226 & \$357 \\
Universal distillation (Sec.~4.5)            &     578 &  31{,}649 & \$615 \\
Planted-defect probe (\S\ref{app:planted})   &     701 &  50{,}571 & --- \\
Tool ablation (\S\ref{app:ftdup})            &     324 &  14{,}718 & \$192 \\
\midrule
\textbf{Total}                               & \textbf{15{,}054} & \textbf{737{,}188} & \textbf{\$3{,}667} \\
\bottomrule
\end{tabular}
\caption{Compute accounting for all experiments reported in this paper.}
\label{tab:compute}
\end{table*}

In total the study executed roughly $15$k agent episodes comprising $747k$ agent steps
and a comparable number of model calls, at a mean of $41.3$ steps per rollout. Within the main grid
the split between the two phases is $4{,}233$ rollouts spent inside the evolution loop and
$2{,}820$ on held-out evaluation, i.e.\ evaluation is roughly as expensive as evolution---a consequence of scoring three arms at $k=3$ on every held-out instance of every cell.

The rollout budget is dominated by the grid rather than by any single analysis: the $8\times3$ main
grid plus its mini-SWE-agent reference accounts for $57.5\%$ of all rollouts. The transfer study is
the second largest item because it is quadratic in languages---each ordered pair requires a full
held-out evaluation of the target---which is the practical reason Sec.~4.6 reports a $5\times5$
subset rather than the full matrix in the main text.

\subsection{Cost Accounting}
\label{app:compute-cost}

Costs decompose into two channels. \emph{Rollout cost} is the frozen task policy $\pi$ acting on
instances; \emph{meta-model cost} is the outer loop---the rollout analyzer reading trajectories and
the evolve agent writing edits. Metered meta-model spend across all experiment logs is \$$256$,
against \$$3{,}667$ of metered rollout spend, so the outer loop is roughly $5\%$ of the total. This
ratio is the practical argument for the approach: the expensive part of self-evolution is not
thinking about the harness; it is running the agent often enough to tell whether an edit helped.

\paragraph{Cost per unit of result.} At the observed mean of $41.3$ steps per rollout, one evolved
harness---three evolution rounds on $20$ instances plus a $k=3$ held-out evaluation---costs on the
order of a few hundred rollouts, or single-digit dollars of meta-model spend per cell. This is the
sense in which harness evolution is cheap relative to weight-space adaptation: no gradient step is
taken, the policy is never loaded for training, and the artifact produced is a few kilobytes of
text.

\subsection{Infrastructure}
\label{app:compute-infra}

All experiments ran on a single machine: $64$ CPU cores and $251$~GB of RAM, with no GPU, since
every model is accessed as a remote API and the local work is container orchestration rather than
inference. Each rollout executes inside the benchmark's own per-repository Docker image, and
grading runs in the same image, so the environment a patch is written in is the environment it is
scored in.


\end{document}